\documentclass[pdflatex,sn-mathphys-num]{sn-jnl}

\usepackage{graphicx}%
\usepackage{multirow}%
\usepackage{amsmath,amssymb,amsfonts}%
\usepackage{amsthm}%
\usepackage{mathrsfs}%
\usepackage[title]{appendix}%
\usepackage{xcolor}%
\usepackage{textcomp}%
\usepackage{manyfoot}%
\usepackage{booktabs}%
\usepackage{algorithm}%
\usepackage{algorithmicx}%
\usepackage{algpseudocode}%
\usepackage{listings}%
\usepackage{docmute}
\usepackage{placeins}

\usepackage{caption} 
\theoremstyle{thmstyleone}%

\begin{document}

\title{Collective cooperation without individual fidelity in LLM agents}


\author[1,2]{\fnm{Henrique} \sur{Ferraz de Arruda}}
\author[3]{\fnm{Carlos} \sur{Gracia Lázaro}}
\author[2,4]{\fnm{Alberto} \sur{Aleta}}
\author[2,4]{\fnm{Yamir} \sur{Moreno}}

\affil[1]{\orgname{ARAID Foundation}, \orgaddress{\city{Zaragoza}, \country{Spain}}}
\affil[2]{\orgdiv{Institute for Biocomputation and Physics of Complex Systems}, \orgname{University of Zaragoza}, \orgaddress{\city{Zaragoza}, \country{Spain}}}
\affil[3]{\orgname{Universidad San Jorge (USJ)}, \orgaddress{\city{Zaragoza}, \country{Spain}}}
\affil[4]{\orgdiv{Department of Theoretical Physics}, \orgname{University of Zaragoza}, \orgaddress{\city{Zaragoza}, \country{Spain}}}


\abstract{
Large language models (LLMs) are increasingly used as agents in simulations of social systems, yet it remains unclear when their behavior can be interpreted as a faithful proxy for human decision-making. Here we test LLM agents against a direct empirical benchmark: a large-scale networked Prisoner's Dilemma experiment with human participants. Using the same interaction protocol, payoff structure, and network topologies, we compare nine open-weight LLMs with the human data. The selected model reproduces several macro-level features of cooperation dynamics, including the early decline and later stabilization of cooperation. This aggregate agreement, however, does not extend uniformly to finer levels of behavior. LLM populations underestimate individual-level heterogeneity and generate conditional cooperation patterns that differ from those observed in humans. Adding a fraction of random agents improves some aspects of micro-level agreement, but does not remove the mismatch in decision rules. These findings reveal a macro--micro dissociation in LLM-based social agents: collective outcomes can appear human-like even when the underlying behavioral distributions and mechanisms are not. They suggest that validating LLM agents as human surrogates requires comparisons across aggregate dynamics, individual heterogeneity, and context-dependent decision rules, rather than outcome-level agreement alone.
}





\maketitle

\section{\label{sec:introduction}Introduction}

Large language models (LLMs) are increasingly used as proxies for human agents in simulations, evaluation settings, and collective decision-making environments~\cite{lu2024llms, chiang2023can, papachristou2025leveraging}. This development is part of a broader shift in which language models are no longer treated only as systems that generate text, but also as tools to build agents whose behavior can be elicited, measured, and compared with that of humans. Recent work has extended this approach to experimental and agent-based settings, where LLM agents are used to simulate human-like behavior in social and strategic contexts~\cite{lu2024llms, flamino2025testing}.

The appeal of this approach is clear. If LLM agents could reproduce human behavior at scale, they would provide a powerful tool for computational social science, behavioral experimentation, and the construction of social simulations. Yet this promise also raises a basic validation problem. Agreement with human outputs does not necessarily imply that a model reproduces the behavioral patterns or decision processes that generated those outputs. Several recent studies point in this direction. LLMs may reconstruct users through compressed behavioral representations that introduce systematic biases in social simulations~\cite{nudo2026generative}; they may align with human evaluative judgments while relying more heavily on lexical associations and statistical priors than on contextual reasoning~\cite{loru2025simulation}; and LLM-generated text can display statistical signatures, such as higher structural regularity and compressibility, that differ from human language production~\cite{hadad2026statistical}. These observations suggest that apparent behavioral realism may be shallow if it is assessed only at the level of visible outcomes.

This issue is especially important in social simulations and game-theoretic environments, where LLM agents are increasingly used as substitutes for human participants. LLMs can participate effectively in repeated strategic interactions, particularly in self-interested settings such as the iterated Prisoner's Dilemma family, while showing weaker performance in coordination problems~\cite{akata2025playing}. Other work has shown that LLM agents can develop collective conventions and shared behavioral patterns without explicit programming~\cite{ashery2025emergent}. At the same time, these agents remain distinguishable from humans in interactive settings: in online debates, for example, LLM agents can remain on topic and blend into conversations, while still being perceived as less convincing and less confident than human interlocutors~\cite{flamino2025testing}. Related studies show that LLM outputs can approach human benchmarks in some evaluative tasks and can be systematically shaped through prompting and fine-tuning~\cite{kumar2026large, serapio2025psychometric}.

A further complication is that current LLMs are not primarily optimized to reproduce human behavior in experiments. They are generally trained and aligned to function as helpful assistants in human-oriented tasks~\cite{wang2023aligning}. This distinction matters because a model that is useful, compliant, or socially desirable need not be a faithful model of human decision-making. Alignment procedures may also compress heterogeneous preferences into narrower sets of high-probability behaviors, as shown by Slocum et al.~\cite{slocum2025diverse}, with related empirical evidence of mode collapse in aligned language models~\cite{hamilton2024detecting}. These concerns are consistent with recent commentaries emphasizing the lack of empirical validation against real human behavioral data in LLM-based agent models~\cite{reia2025opportunities}, and with calls to calibrate scientific claims from LLM social simulations to the strength of robustness checks and empirical benchmarks~\cite{ye2026stop}.

Here, we evaluate whether LLM agents can function as behavioral surrogates for humans in repeated networked social dilemmas. We use as a benchmark the large-scale Prisoner's Dilemma experiment of Gracia-Lázaro et al.~\cite{gracia2012heterogeneous}, in which human cooperation converged to similar levels across network topologies, challenging the theoretical expectation that heterogeneous networks should promote cooperation. Using the same network structures and interaction protocol, we compare nine open-weight LLMs operating as autonomous agents in repeated Prisoner's Dilemma games. Our aim is not to ask whether LLMs play the game better than humans, but whether they reproduce the empirical temporal dynamics, individual heterogeneity, and conditional behavioral patterns observed in the experiment.

This design allows us to test behavioral similarity at several levels. At the macro level, we ask whether LLM agents reproduce the aggregate evolution of cooperation. At the micro level, we ask whether they reproduce the distribution of individual cooperation propensities. At the decision-rule level, we ask whether they reproduce conditional cooperation, namely, the way in which participants adjust their actions to the previous behavior of their neighbors. We find that the selected LLM reproduces several macro-level regularities, including the early decay and later stabilization of cooperation. However, this aggregate agreement does not fully extend to micro-level signatures: simulated populations underestimate behavioral heterogeneity and generate conditional cooperation rules that differ quantitatively from the empirical data. These results reveal a macro--micro dissociation in LLM-based social agents, with direct implications for the validation of machine behavior and for the use of LLM agents as human surrogates in experimental social science.

\section{\label{sec:results}Results}

We present the results in three steps. We first describe the empirical benchmark and the simulation framework. We then assess whether LLM agents reproduce aggregate cooperation dynamics, before turning to individual heterogeneity and conditional cooperation. This ordering follows the central validation question of the study: whether agreement at the collective level extends to finer behavioral levels.

\subsection{Empirical benchmark and simulation framework}
To evaluate whether LLM agents reproduce human cooperation dynamics, we simulated a repeated multi-agent Prisoner's Dilemma on network structures derived from a human behavioral experiment. The simulation framework consists of an orchestration layer that manages autonomous agents, each controlled by an LLM. 
This work builds on the experimental design of Gracia-Lázaro et al.~\cite{gracia2012heterogeneous}, who examined human cooperation in networked Prisoner's Dilemma games. The original study found that human cooperation converged to similar steady-state levels across network topologies, challenging theoretical expectations that heterogeneous networks should promote cooperation. Using the same network structures and temporal dynamics as in the original study, we test whether LLM-based agents reproduce the empirical cooperation patterns or instead follow behavior closer to classical game-theoretic predictions.

The simulation environment is a repeated multi-agent game in which each network node is occupied by an LLM-driven agent. The simulation proceeds in discrete rounds that mirror the original experiment. In each round, after the introductory one, agents receive a prompt summarizing the previous round, including their neighbors' choices and normalized payoffs. Agents then return their choice for the new one which effectively could be to \emph{cooperate} or to \emph{defect}. The underlying incentive structure is a Prisoner's Dilemma where payoffs are computed pairwise. The total payoff for an agent is the sum of these pairwise interactions across all their network connections. The specific values used to drive the agents' goal of maximizing ``ECUs'' (Experimental Currency Units) are detailed in Table~\ref{tab:payoffs} and follow Grujic et al~(2010)~\cite{grujic2010social} as in the original experiment. In this specific configuration, the rewards are designed such that cooperation is theoretically expected to reach a high level~\cite{gracia2012heterogeneous, grujic2010social}. Implementation details, model descriptions, and statistical procedures are provided in the Methods.

\begin{table}[h]
\centering
\caption{Pairwise payoff matrix (ECUs per neighbor).}
\label{tab:payoffs}
\begin{tabular}{l|cc}
\toprule
\textbf{Your Choice / Neighbor Choice} & \textbf{Cooperate} & \textbf{Defect} \\ \hline
\textbf{Cooperate}                         & 7              & 0              \\
\textbf{Defect}                         & 10             & 0              \\ \bottomrule
\end{tabular}
\end{table}

\subsection{Macro-level cooperation dynamics}

\begin{figure}[ht!]
    \centering
    \includegraphics[width=1.\linewidth]{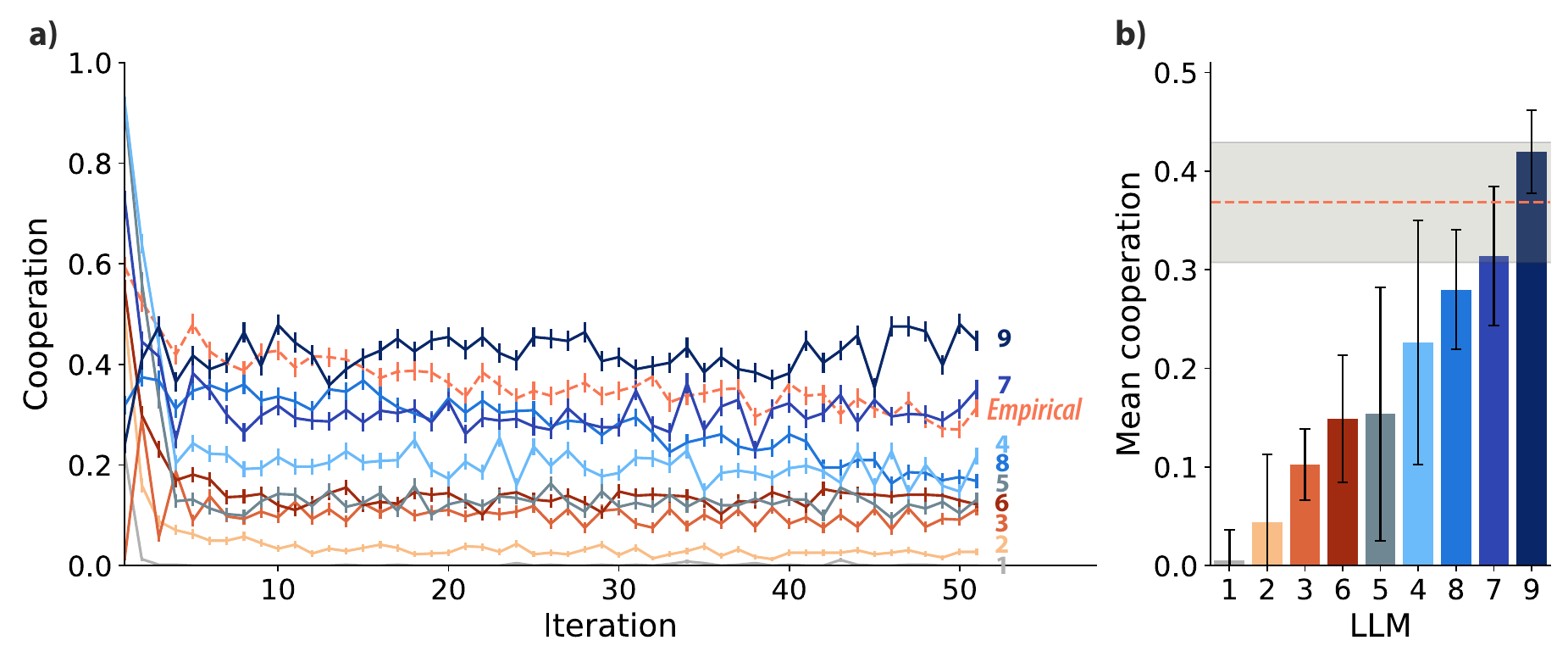}
    \caption{
    \textbf{Comparison of cooperation outcomes across LLM models.} 
    The numbers shown on the plot correspond to the following models:
    (1) BlackHillsInfoSec llama-3.1:8b-abliterated,
    (2) deepseek-llm:67b,
    (3) deepseek-llm:7b,
    (4) krith meta-llama-3.1:70b-instruct-abliterated IQ4 XS,
    (5) llama3.1:70b,
    (6) llama3.1:8b,
    (7) llama4:16x17b,
    (8) smolLM2.1:7b, and
    (9) qwen3:32b.
    Experiments performed for the Lattice network.
    Panel~(a) shows the mean cooperation rate per round (empirical) or per iteration (simulation), and the error bars represent the standard error. Panel~(b) shows the average values of the simulation curves, with error bars representing the standard deviation. The dashed horizontal line represents the empirical average, and the shaded area represents one standard deviation around it.
    }
    \label{fig:full_simulation_comparison}
\end{figure}

We first ask whether different LLMs produce comparable aggregate cooperation dynamics when placed in the same experimental environment. For this initial benchmark, we fix the interaction structure to a regular lattice network (625 nodes, average degree 4). Figure~\ref{fig:full_simulation_comparison}(a) shows the resulting cooperation trajectories across all models, while Fig.~\ref{fig:full_simulation_comparison}(b) shows their average cooperation levels. The tested LLMs span several model families, parameter scales, and training paradigms. Despite identical incentives and network structure, they generate markedly different cooperation regimes. Notably, \emph{qwen3:32b} produces the highest cooperation levels across the simulation horizon. Ablated variants derived from the same base checkpoints also display systematically altered trajectories relative to their originals, indicating that collective outcomes are sensitive to model choice and alignment. Among the models tested, \emph{llama4:16x17b} yields the closest aggregate match to the empirical data. We therefore use this model as the primary agent in the remaining analyses, while treating the initial benchmark as evidence that LLMs should not be regarded as interchangeable behavioral surrogates.

\begin{figure}[h!]
    \centering
    \includegraphics[width=1\linewidth]{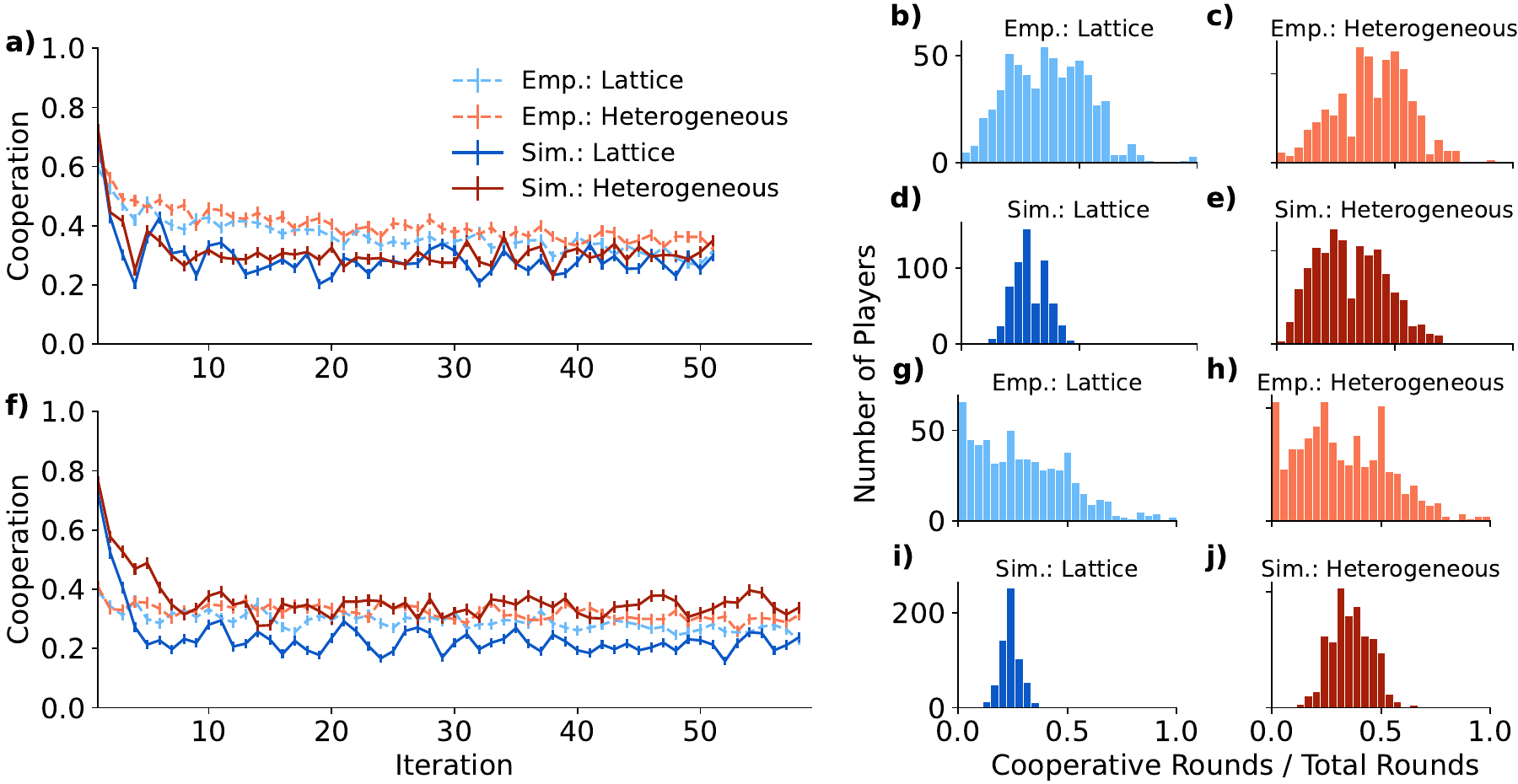}
    \caption{{\textbf{Evolution and distribution of cooperation.}
    Panels~(a) and (f) show the temporal evolution of the mean cooperation rate per iteration for the networked and control experiments, respectively. Dashed lines represent empirical results, solid lines represent simulation results, and error bars indicate the standard error. Panels~(b)-(e) and (g)-(j) display the histograms of individual cooperation propensities (cooperative rounds / total rounds). Specifically, Panels (a)-(e) show the results for the networked experiment, and Panels~(f)-(j) show the results for the control case.}}
    \label{fig:scenarios_comparison}
\end{figure}

With this LLM fixed for the remaining analyses, we next evaluate the two network structures considered in~\cite{gracia2012heterogeneous}: a regular lattice and a heterogeneous network (604 nodes, degrees ranging from $k=2$ to $k=16$). Fig.~\ref{fig:scenarios_comparison}(a) shows the temporal evolution of cooperation for one realization of the dynamics. As in the empirical experiment, both structures yield similar cooperation levels that overlap closely over time. Compared to the empirical data, the simulated trajectories align more closely in later iterations, though they remain noisier overall.

To perform a thorough comparison, we also test the control cases, which consist of simulations on a time-varying network structure. Specifically, connections are randomly rewired at each iteration while preserving the original degree sequence. To match the original study, we do not generate a new rewiring sequence. Instead, we use the exact sequence of network connections reported in~\cite{grujic2010social}, which was originally generated at random. Henceforth, we refer to the fixed-network simulations as the \emph{networked} condition, and to the time-varying simulations as the \emph{control} condition. Figure~\ref{fig:scenarios_comparison}(f) shows the resulting cooperation trajectories over time. Notably, the heterogeneous control case more closely matches the empirical data, whereas in the lattice control case, agents tend to cooperate less. As in the networked simulations, the LLM-based trajectories remain noisier overall.

Upon closer inspection of the cooperation histograms, we observe that the empirical experiments yield substantially broader distributions than those of the simulations under the networked condition. Specifically, the empirical lattice case (Fig.~\ref{fig:scenarios_comparison}b) has mean cooperation $\mu=0.368$ with standard deviation $\sigma=0.170$, and the empirical heterogeneous case (Fig.~\ref{fig:scenarios_comparison}c) has $\mu=0.405$ and $\sigma=0.158$. By contrast, the simulated lattice network (Fig.~\ref{fig:scenarios_comparison}d) produces a more concentrated distribution ($\mu=0.291$, $\sigma=0.069$), whereas the simulated heterogeneous network (Fig.~\ref{fig:scenarios_comparison}e) yields $\mu=0.314$ and $\sigma=0.148$. While the lattice simulation more closely reproduces the empirical mean, the heterogeneous simulation better captures the observed dispersion and therefore provides a closer overall match. 

To test the sensitivity of our results to agent variability, we repeated the lattice simulation with LLM temperatures sampled uniformly at random. The resulting distributions were similar to those reported in the main text, showing that our findings are not affected by temperature variation (see Supplementary Material, Fig.~S13).  We also test three different ``personas'' for the LLMs. However, since the results are consistent across personas, we omitted them from the paper (see Supplementary Material, Section~S3.2.1, for more details). Furthermore, to evaluate whether the reduced dispersion observed in the simulated cooperation distributions is specific to \texttt{llama4\_16x17b}, we repeated the analysis using \texttt{qwen3:32b}, the second-best-performing model according to Fig.~\ref{fig:full_simulation_comparison}. As shown in Supplementary Fig.~S14, \texttt{qwen3:32b} reproduces the same qualitative pattern, in which the simulated cooperation distributions remain narrower than the corresponding empirical distributions. These results suggest that the tendency of LLM-based agents to generate less variable outcomes than human participants is not restricted to a single model.

A similar pattern is observed in the \emph{control} condition. The empirical lattice control (Fig.~\ref{fig:scenarios_comparison}g) exhibits $\mu=0.292$ and $\sigma=0.209$, and the empirical heterogeneous control (Fig.~\ref{fig:scenarios_comparison}h) shows $\mu=0.322$ and $\sigma=0.209$, again indicating wide distributions. In contrast, the simulated lattice control (Fig.~\ref{fig:scenarios_comparison}i) remains narrowly concentrated ($\mu=0.238$, $\sigma=0.047$), while the simulated heterogeneous control (Fig.~\ref{fig:scenarios_comparison}j) produces higher cooperation ($\mu=0.361$) but still substantially reduced variability ($\sigma=0.089$). Thus, although the heterogeneous control simulation better matches the empirical mean cooperation level, both simulated control cases underestimate the dispersion observed in human behavior.

To evaluate the fidelity of the simulated temporal dynamics, we used metrics to compare simulated and empirical cooperation trajectories. In addition to measuring point-by-point error with Root Mean Squared Error (RMSE) and Mean Absolute Error (MAE), we used an autoregressive model (AR(1)) to evaluate the persistence of cooperation. The difference in persistence coefficients ($\text{Diff}_\phi$) indicates whether simulated and empirical trajectories exhibit similar temporal persistence (i.e., a form of aggregate ``memory'' in the cooperation series). The difference in implied mean increments ($\text{Diff}_\mu$) identifies systematic biases in the average rate of change. Finally, to measure the synchrony of cooperation trends, we calculated the Pearson correlation coefficient for the raw levels ($r$) and the first differences ($r^\Delta$), ensuring that the simulated fluctuations were synchronized with the empirical observations. To ensure that these results were not due to random chance, we established statistical significance using a Monte Carlo null model that preserves marginal distributions while removing temporal structure.

Table~\ref{tab:Errors_types} shows that the simulations reproduce the overall shape of the empirical cooperation trajectories, but still deviate systematically in temporal structure. Pointwise errors (RMSE and MAE) are lowest in the \emph{heterogeneous control} condition and highest in the \emph{heterogeneous} networked condition. The AR(1) persistence comparison shows that lattice simulations differ strongly from humans in decision ``memory,'' with large and significant $\text{Diff}{_\phi}$ values (0.217 and 0.464), whereas the heterogeneous networked case shows the closest match in persistence ($\text{Diff}{_\phi}=0.071$). Differences in implied long-run means remain small across all conditions ($\text{Diff}_{\mu}\leq 0.004$), suggesting limited bias in average cooperation levels. Finally, correlations in raw cooperation levels are consistently moderate and significant, but correlations in first differences are weaker, indicating that the simulations capture broad trends better than round-to-round fluctuations.

\begin{table}[h!]
    \centering
    \scriptsize
    \begin{tabular}{lllllll}
    \toprule
    Experiment & RMSE & MAE & $\text{Diff}_\phi$ & $\text{Diff}_\mu$ & $r$ & $r^{\Delta}$ \\
    \midrule
    lattice & 0.098** & 0.083** & 0.217*** & 0.002 & 0.612*** & 0.276* \\
    lattice control & 0.089*** & 0.074* & 0.464*** & 0.002*** & 0.583*** & 0.214 \\
    heterogeneous & 0.107** & 0.096 & 0.071 & 0.002 & 0.622*** & 0.151 \\
    heterogeneous control & 0.079*** & 0.051 & 0.359 & 0.004* & 0.440*** & 0.362*** \\
    \bottomrule
    \end{tabular}

    \caption{\textbf{Trajectory errors, autoregressive comparison, and correlation}. RMSE and MAE measure point-by-point deviations between empirical and simulated cooperation rates. $\text{Diff}_{\phi}$ denotes the absolute difference between the AR(1) persistence coefficients, and $\text{Diff}_{\mu}$ denotes the absolute difference between the implied long-run means of the simulated and empirical series. $r$ denotes the Pearson correlation coefficient between empirical and simulated trajectories. $r^\Delta$ represents the Pearson correlation coefficient using the first differences. For RMSE, MAE, $\text{Diff}_{\phi}$, and $\text{Diff}_{\mu}$, statistical significance is assessed using the Monte Carlo null model described in the text. For $r$ and $r^\Delta$, significance refers to the null hypothesis of zero correlation ($H_0: r = 0$). $^{***}$ denotes $p < 0.01$, $^{**}$ denotes $p < 0.05$, $^{*}$ denotes $p < 0.1$, and no symbol indicates that the result is not statistically significant.}
    \label{tab:Errors_types}
\end{table}

As an exploratory test of whether non-strategic behavior contributes to human-like variability, we introduced a fraction of random agents. This choice is motivated by qualitative evidence from the original experiment suggesting that some participants may have played without a stable strategy, or even at random~\cite{gracia2012heterogeneous}. To reproduce this situation, we assign a fraction $\rho$ of nodes to act as \emph{random agents}. These agents do not follow the LLM policy, but choose between cooperate and defect uniformly at random in every round. The remaining fraction $(1-\rho)$ of nodes is controlled by the LLM-driven decision rule. Based on a systematic sweep over values of $\rho$ (Supplementary Information, Table~S3), we use $\rho=0.2$ as an illustrative hybrid condition. This value balances improved temporal alignment with the empirical cooperation trajectories, particularly in terms of fluctuation synchrony ($r^{\Delta}=0.469$), with preservation of the overall trajectory shape ($r=0.690$), while avoiding the stronger distortions observed at higher random-agent fractions. Alternative trajectories for other values of $\rho$ are shown in Supplementary Information, Fig.~S15.

\begin{figure}[h!]
    \centering
    \includegraphics[width=1.\linewidth]{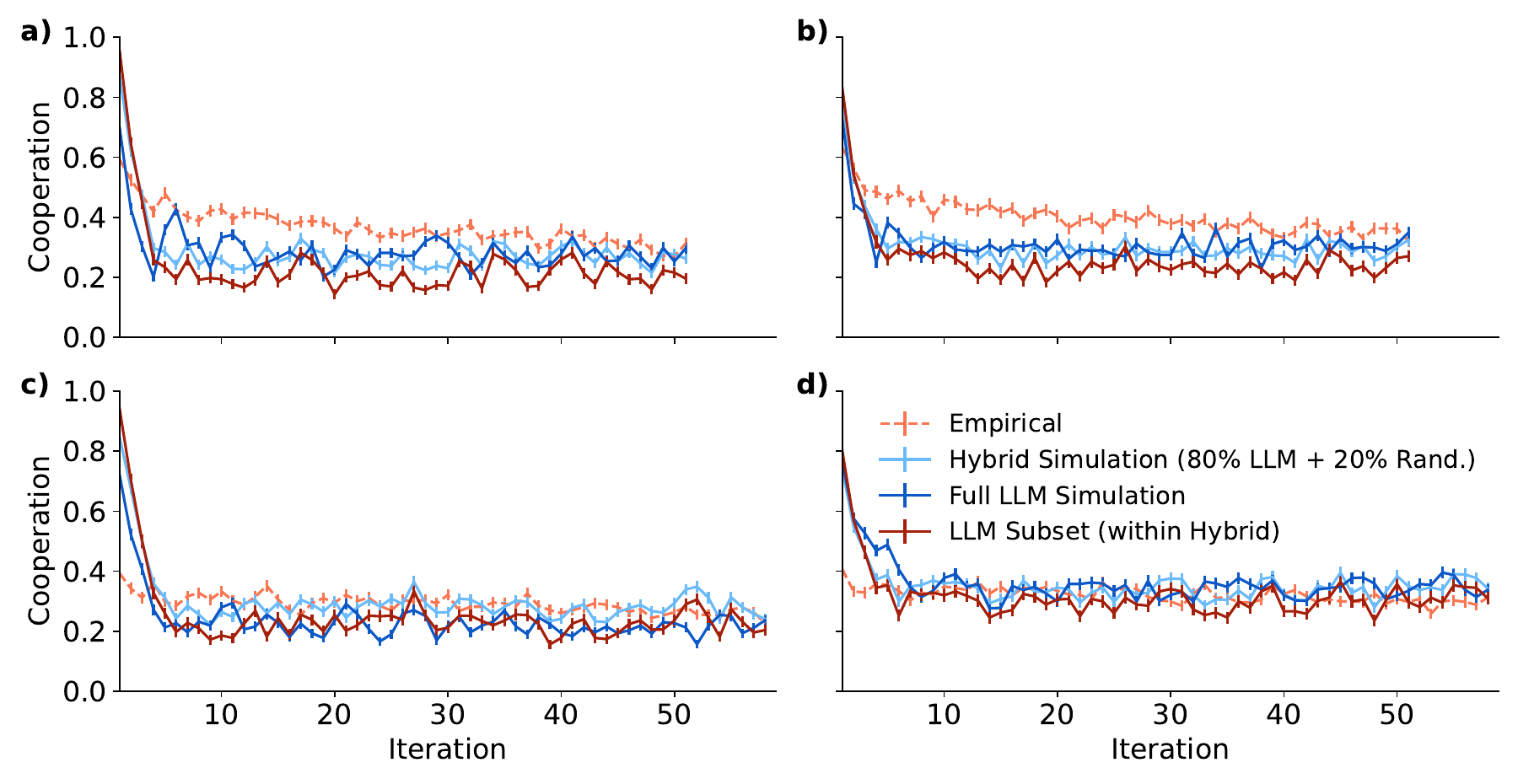}
    \caption{\textbf{Evolution of cooperation with a fraction of random agents ($\rho=0.2$)}. Panels show cooperation over iterations for (a)~\texttt{lattice}, (b)~\texttt{lattice control}, (c)~\texttt{heterogeneous}, and (d)~\texttt{heterogeneous control} network structures. Solid lines denote the mean cooperation rate per round (empirical) or per iteration (simulation), and error bars represent the standard error.}
    \label{fig:20_with_rand}
\end{figure}

Figure~\ref{fig:20_with_rand} shows that introducing a fraction of random agents ($\rho=0.2$) preserves the overall qualitative shape of the empirical trajectories, with cooperation rapidly decaying in early rounds and subsequently stabilizing. However, the inclusion of random agents alters the quantitative agreement with human data in a condition-dependent manner. In the lattice network, adding random agents increases the correlation with the empirical trajectory and substantially improves synchrony in short-term fluctuations ($r^{\Delta}$), although pointwise errors increase slightly. A similar pattern holds for the heterogeneous network, where the hybrid case yields stronger correlation ($r=0.743$ vs. $r=0.622$) and higher agreement in first differences ($r^{\Delta}=0.343$ vs. $0.151$), while maintaining comparable error levels (RMSE $=0.109$ vs. $0.107$). 

In the control scenarios, the effect of random agents is more mixed. In the lattice control case, the hybrid simulation slightly reduces RMSE (0.087 vs. 0.089) and markedly lowers MAE (0.046 vs. 0.074), but decreases correlation with the empirical trajectory ($r=0.509$ vs. $r=0.583$). In the heterogeneous control case, the hybrid configuration yields the lowest RMSE across all conditions (0.070), but also produces a weaker correlation ($r=0.325$), compared to $r=0.440$ in the scenario without random agents. See Supplementary Material, Table~S4, for more details on all the trajectory errors. Together, these results indicate that random agents can improve alignment with human temporal fluctuations in the networked condition, while in the control condition, they primarily reduce absolute error without consistently improving temporal synchrony.

\subsection{Micro-level heterogeneity and decision rules}
We next ask whether aggregate agreement is accompanied by individual-level fidelity. To do so, we compare the distribution of cooperation propensities across agents, rather than only the population mean. This distinction is important because two populations can display similar average cooperation while differing substantially in the diversity of individual behaviors. We quantify the distance between empirical and simulated propensity distributions using the first-order Wasserstein distance, $W_1(E,S)$. Lower values indicate closer agreement in behavioral heterogeneity, whereas larger values indicate that the simulated population differs from the empirical one in how cooperation is distributed across individuals. Table~\ref{tab:heterogeneity_results_main} summarizes the results across experimental scenarios.

Table~\ref{tab:heterogeneity_results_main} shows that the simulations match empirical individual-level heterogeneity with varying accuracy across experimental scenarios. The smallest distance is obtained in the lattice network with mixed populations, where $W_1(E,S)=0.081$, indicating the closest match to the empirical distribution of cooperation propensities. In contrast, the lattice control condition with a fully LLM-driven population yields the largest discrepancy,  $W_1(E,S)=0.136$. Introducing random agents generally reduces the distance relative to the corresponding fully LLM-driven case, with the strongest improvement observed in the lattice control condition (from $0.136$ to $0.095$). In the heterogeneous network, however, the mixed-population case slightly worsens the match under the networked condition (from $0.091$ to $0.095$).

\begin{table}[h!]
\centering
\begin{tabular}{lll}
    \toprule
    Experiment & $W_1(E,S)$ & $W_1(E, Null) $\\
    \midrule
    lattice (100\% LLM + 0\% Rand.) & 0.101 & 0.085 [95\% CI: 0.082, 0.089] \\
    lattice (80\% LLM + 20\% Rand.) & 0.081 & 0.085 [95\% CI: 0.082, 0.089] \\
    lattice control (100\% LLM + 0\% Rand.) & 0.136 & 0.125 [95\% CI: 0.123, 0.128]\\
    lattice control (80\% LLM + 20\% Rand.) & 0.095 & 0.125 [95\% CI: 0.123, 0.128]\\
    heterogeneous (100\% LLM + 0\% Rand.) & 0.091 & 0.074 [95\% CI: 0.071, 0.077] \\
    heterogeneous (80\% LLM + 20\% Rand.) & 0.095 & 0.074 [95\% CI: 0.071, 0.077] \\
    heterogeneous control (100\% LLM + 0\% Rand.) & 0.109 & 0.126 [95\% CI: 0.123, 0.129] \\
    heterogeneous control (80\% LLM + 20\% Rand.) & 0.093 & 0.126 [95\% CI: 0.123, 0.129] \\
    \bottomrule
    \end{tabular}
\caption{\textbf{Wasserstein distance between empirical and simulated distributions of individual cooperation propensities across experimental scenarios.} Lower values indicate better reproduction of behavioral heterogeneity. The null model assumes that each agent follows an independent, identically distributed, Bernoulli process with the network's empirical mean cooperation rate, all while maintaining the same number of rounds per agent.}
\label{tab:heterogeneity_results_main}
\end{table}

For each network condition (lattice, lattice control, heterogeneous, heterogeneous control), we quantified behavioral heterogeneity in the empirical data as the distribution of agent-level cooperation propensities (fraction of cooperative actions per agent). We then generated a parametric null model in which each agent follows an independent Bernoulli process with success probability equal to the network-level empirical mean cooperation rate, while preserving each agent's number of rounds. Across (B=10{,}000) null realizations, we computed $W_1(E,\mathrm{Null})$ between empirical and null propensity distributions. We report the mean null distance and a 95\% confidence interval of this Monte Carlo distribution, providing a baseline for whether observed heterogeneity departs from what is expected under homogeneous random behavior.

To contextualize these values, we compare the results of Table~\ref{tab:heterogeneity_results_main} with the null-model baseline. In the lattice network, the null model yields $W_1(E,\mathrm{Null})=0.085$ (95\% CI: [0.082, 0.089]), which is close to the best-performing simulation ($W_1(E,S)=0.081$), indicating that the empirical heterogeneity in this condition can largely be explained by variation consistent with independent behavior around the population mean. In the lattice control condition $W_1(E,\mathrm{Null})=0.125$ (95\% CI: [0.123, 0.128]), which is not comparable to the full-LLM case ($W_1(E,S)=0.136$), while the mixed simulation ($W_1(E,S)=0.095$) falls below the null baseline. In the heterogeneous network, the null distance is smaller ($W_1(E,\mathrm{Null})=0.074$, 95\% CI: [0.071, 0.077]) than both simulation cases ($W_1(E,S)=0.091$ and $0.095$), suggesting that the simulated populations exhibit less empirical-like heterogeneity than expected under the null benchmark. Finally, in the heterogeneous control condition, the null distance is again high ($W_1(E,\mathrm{Null})=0.126$, 95\% CI: [0.123, 0.129]), whereas both simulation scenarios yield lower distances ($W_1(E,S)=0.109$ and $0.093$), indicating improved agreement with empirical heterogeneity compared to the null expectation.

We then examine a stricter behavioral signature: \emph{conditional cooperation}. This measures the probability that an agent cooperates as a function of the fraction of cooperating neighbors in the previous round, and according to whether the agent cooperated or defected in the previous round. Conditional cooperation is important because it probes the local decision rule, rather than only the resulting level of cooperation. As in the empirical study, we first analyze the networked condition, where neighborhoods remain fixed. In the human experiment, this relationship is well approximated by a linear trend with distinct responses following cooperation and defection. Across most simulated scenarios, we observe a qualitative dependence on neighborhood cooperation, but the responses are quantitatively different and often steeper than in the human data. The ``after defection'' response deviates most strongly from the empirical pattern in the lattice network, exhibiting an inverted slope in both the full-LLM and hybrid simulations. Moreover, unlike the empirical data, the simulated conditional-cooperation points do not consistently follow an approximately linear relationship, making the linear-fit comparison used in the original study less informative here. We therefore report the full curves in the Supplementary Material, Section~S3.2.2, highlighting an additional discrepancy between LLM-based agents and empirical behavioral patterns.

\begin{figure}[h!]
\centering
\includegraphics[width=1\linewidth]{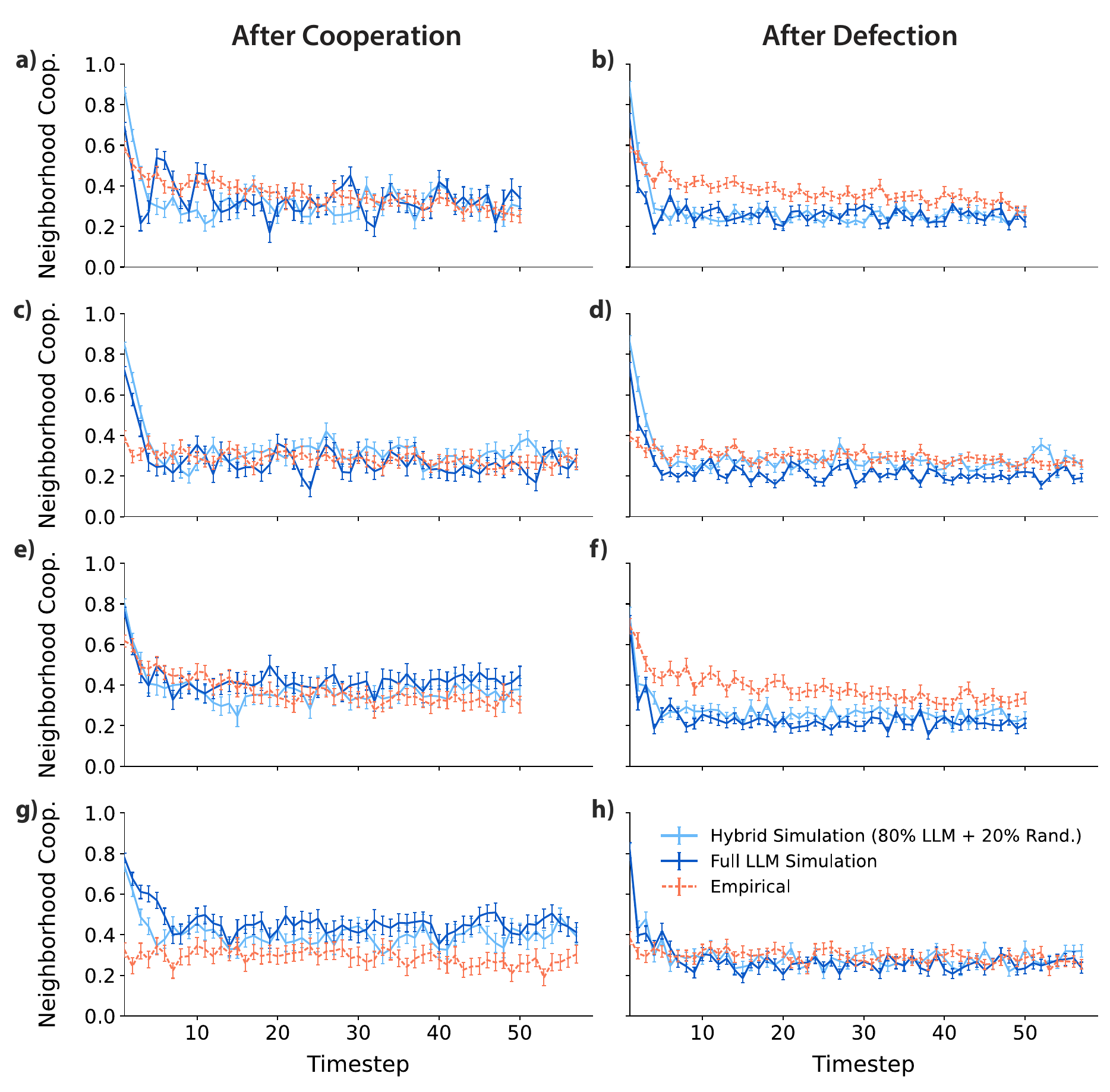}
\caption{\textbf{Conditional cooperation across network structures and experimental conditions.} Each panel shows the estimated probability of cooperation as a function of the fraction of cooperating neighbors in the previous round, computed from a single execution of the dynamics. Panels~(a) and (b) correspond to the lattice network under the networked condition, (c) and (d) to the lattice network under the control condition, (e) and (f) to the heterogeneous network under the networked condition, and (g) and (h) to the heterogeneous network under the control condition. In each case, empirical conditional cooperation patterns are compared with simulations under different agent compositions, including fully LLM-driven and mixed populations with random agents. Error bars indicate 95\% confidence intervals.}
\label{fig:neighborhood_cooperation}
\end{figure}

To extend this analysis, we examine conditional cooperation across network structures and experimental conditions. Figure~\ref{fig:neighborhood_cooperation} shows conditional cooperation across experimental scenarios for a single execution of the dynamics. In each case, the curves represent the estimated conditional probabilities of cooperation as a function of neighborhood cooperation in the previous round. We repeated this conditional-cooperation analysis over 20 independent executions of the dynamics and obtained consistent curves with narrow confidence intervals, indicating that a single execution is sufficient to capture the qualitative conditional-cooperation trends of the simulations (for details see Supplementary Material, Figs.~S22~and~S23).

The quantitative deviations reported in Table~\ref{tab:table_conditional_rule_all_experiments} are consistent with the qualitative patterns shown in Fig.~\ref{fig:neighborhood_cooperation}. Across most scenarios, errors are larger in the $z=1$ case (after cooperation) than in the $z=0$ case (after defection), reflecting stronger discrepancies in how simulated agents condition their behavior on neighborhood cooperation following a cooperative move. The largest deviations are observed in the lattice network under the networked condition, where the full-LLM population yields RMSE$_{z=1}=0.397$ and MAE$_{z=1}=0.388$, matching the visibly distorted and non-linear response in panel A. Introducing random agents reduces these discrepancies in the lattice networked case (RMSE$_{z=1}$ decreases to 0.255), although a substantial mismatch remains.

In contrast, the lattice control condition shows comparatively small error after defection (RMSE$_{z=0}=0.046$ in the full-LLM case), consistent with the flatter conditional patterns visible under rewiring, while errors after cooperation remain large (RMSE$_{z=1}=0.341$). For the heterogeneous network, the mixed population reduces deviations relative to the full-LLM case, particularly in the $z=1$ regime (RMSE$_{z=1}$ decreases from 0.355 to 0.232), which aligns with the closer agreement between the empirical and simulated curves observed in the corresponding panels. Overall, these results confirm that while simulations capture the qualitative dependence of cooperation on local neighborhood behavior, the inferred conditional rules remain quantitatively distinct from the empirical patterns, particularly following cooperative actions.

\begin{table}[h!]
\centering
\begin{tabular}{lllll}
\toprule
Experiment & RMSE$_{z=0}$ & MAE$_{z=0}$ & RMSE$_{z=1}$ & MAE$_{z=1}$ \\
\midrule
lattice (100\% LLM + 0\% Rand.) & 0.237** & 0.217*** & 0.397*** & 0.388* \\
lattice (80\% LLM + 20\% Rand.) & 0.180 & 0.165*** & 0.255*** & 0.246*\\
lattice control (100\% LLM + 0\% Rand.) & 0.046*** & 0.038*** & 0.341*** & 0.337***\\
lattice control (80\% LLM + 20\% Rand.) & 0.090* & 0.086* & 0.221*** & 0.211*** \\
heterogeneous (100\% LLM + 0\% Rand.) & 0.177*** & 0.154 & 0.355*** & 0.320*** \\
heterogeneous (80\% LLM + 20\% Rand.) & 0.161*** & 0.140 & 0.232** & 0.214 \\
heterogeneous control (100\% LLM + 0\% Rand.) & 0.107** & 0.091*** & 0.217 & 0.196 \\
heterogeneous control (80\% LLM + 20\% Rand.) & 0.190 & 0.124 & 0.173 & 0.161**\\
\bottomrule
\end{tabular}
\caption{\textbf{Deviation between empirical and simulated conditional cooperation rules across experimental scenarios.}
Errors are reported separately for cases in which the previous action was defection ($z=0$) or cooperation ($z=1$). $^{***}$ denotes $p < 0.01$, $^{**}$ denotes $p < 0.05$, $^{*}$ denotes $p < 0.1$, and no symbol indicates that the result is not statistically significant.}
\label{tab:table_conditional_rule_all_experiments}
\end{table}

\section{\label{sec:discussion}Discussion}

This study tested whether autonomous LLM agents can serve as behavioral surrogates for humans in a repeated networked social dilemma. The answer is mixed in a revealing way. The selected model, \emph{llama4:16x17b}, reproduces several macro-level features of the human experiment, including the early decline and later stabilization of cooperation observed by Gracia-Lázaro et al.~\cite{gracia2012heterogeneous}, as well as the absence of network reciprocity reported in the original study. Yet this aggregate agreement does not imply individual-level fidelity. The simulated populations underestimate important aspects of human heterogeneity and generate conditional cooperation patterns that differ from the empirical decision rules.

This pattern is the central result of the paper. LLM-based agents can reproduce collective cooperation dynamics without fully reproducing the distribution of individual behaviors or the context-dependent mechanisms through which humans adapt their decisions. In other words, the model captures part of the phenomenology of the experiment, but not all of its behavioral structure. This macro--micro dissociation matters because many proposed uses of LLM agents in social simulation rely implicitly on the assumption that matching aggregate outcomes is sufficient evidence of behavioral realism.

Our model benchmark also shows that LLMs cannot be treated as interchangeable agents. Different model families produce markedly different cooperation regimes, and alignment-perturbed variants derived from related checkpoints can generate systematically different collective dynamics. Model choice is therefore a substantive modeling assumption, not a technical detail. This sensitivity reinforces the need for empirical benchmarking before drawing conclusions from LLM-based social simulations~\cite{reia2025opportunities}.

The micro-level results are especially informative. Human participants display broad distributions of cooperation propensities, whereas the simulated populations often produce narrower or differently shaped distributions. Conditional cooperation provides an even stronger test. Human behavior in the original experiment is characterized by structured responses to the local social environment, with distinct patterns after cooperation and after defection. The LLM agents reproduce some qualitative dependence on neighborhood cooperation, but the inferred rules differ quantitatively, and in some cases qualitatively, from the empirical patterns. Thus, the model does not simply implement a noisy version of the human rule. It appears to reach partially similar aggregate outcomes through different local response functions.

The hybrid simulations with random agents help clarify this point. Introducing a fraction of random decision-makers improves agreement with the empirical data in some dimensions, particularly fluctuation synchrony in the networked condition and distributional agreement in selected cases. This should not be interpreted as a tuned correction that makes the model human-like. Rather, it suggests that stochasticity, inconsistent behavior, or non-strategic play may be necessary ingredients in models of human cooperation. Real participants are not homogeneous optimizers, and a faithful behavioral surrogate may need to represent this diversity explicitly. At the same time, the hybrid model does not remove the mismatch in conditional cooperation, indicating that noise alone is not sufficient to recover human decision mechanisms.

These results have broader implications for the validation of machine behavior. A model may be useful for reproducing an aggregate trajectory while still being inadequate for causal explanation or intervention design. This distinction is particularly important for digital twins, policy simulations, and agent-based models that use LLMs as substitutes for human actors. If the aim is only to forecast a limited aggregate outcome under conditions close to the benchmark, macro-level agreement may be informative. If the aim is to understand how individuals respond to incentives, social context, or interventions, then micro-level heterogeneity and decision-level rules become essential.

Several limitations should be kept in mind. The empirical benchmark comes from a specific experiment, participant pool, and institutional context, and the same comparison should be repeated across other social dilemmas and populations. The simulations also rely on a particular prompt structure and on open-weight models available at the time of the study. Although persona variants and repeated executions did not qualitatively change the results, broader prompt and model robustness remain an important direction for future work. More broadly, our findings suggest that LLM simulations cannot currently replace human experimentation, which remains necessary to establish the empirical ground truth used for model validation. However, when benchmarked against human data, LLM agents may still be valuable for scaling simulations and exploring conditions that are difficult to study experimentally. Finally, current LLMs are typically optimized to follow human instructions and preferences, not to reproduce the full diversity of human behavior in experiments~\cite{wang2023aligning}. This may contribute to the reduced behavioral diversity observed here, consistent with forms of model homogenization and mode collapse reported in recent work~\cite{slocum2025diverse, hamilton2024detecting}. More generally, our findings support the view that claims based on LLM social simulations should be calibrated to the strength of empirical validation and robustness checks~\cite{ye2026stop}.

Overall, \emph{llama4:16x17b} provides a promising but incomplete surrogate for human behavior in networked social dilemmas. It reproduces robust macro-level cooperation dynamics, but not the full heterogeneity and conditional structure of human strategic decision-making. The main lesson is therefore not simply that LLM agents succeed or fail as human surrogates, but that surrogacy is level-dependent. Validating machine behavior requires asking not only whether artificial agents produce similar outcomes, but also whether they do so through similar behavioral distributions and decision mechanisms.

\section{\label{sec:methods}Methods}

\subsection{Simulation framework}
All LLM-based agents were implemented using the CrewAI orchestration framework\footnote{\url{https://www.crewai.com/}, accessed March 2026}, which handles prompt construction, model querying, response parsing, and payoff updates at each iteration. Model inference was performed locally using Ollama\footnote{\url{https://ollama.com/}, accessed March 2026} as the serving backend, enabling self-contained execution of all agents without external API dependencies. Each agent was mapped to a node in the network and interacted only with the information in its local prompt, which contained its neighbors' previous-round actions and the corresponding normalized payoffs. Models were queried independently for each agent at each round, ensuring decentralized decision-making consistent with the experimental protocol.

We simulate a repeated Prisoner's Dilemma game on empirical social networks (using the same network topologies and temporal dynamics as in Gracia-Lázaro \emph{et al.}~\cite{gracia2012heterogeneous}). Each node (agent) chooses between GREEN and BROWN in each round, corresponding to cooperation and defection, respectively. We use color labels rather than the canonical game actions in the agent-facing prompts to reduce the likelihood that the LLM recognizes the task from familiar Prisoner's Dilemma terminology and responds on that basis rather than from the local interaction history and payoffs. The game proceeds in discrete rounds. In each round, each agent observes its neighbors' previous choices and (degree-)normalized payoffs, then chooses an action for the current round.  Payoffs were computed pairwise as follows: if both agents choose GREEN, each receives 7 ECUs; if one chooses BROWN and the neighbor chooses GREEN, the BROWN agent receives 10 ECUs, and the GREEN agent receives 0 ECUs; if both choose BROWN, both receive 0 ECUs.  An agent's total payoff in a round is the sum of these pairwise payoffs with all neighbors. In the last experiments, a fraction $\rho$ of nodes were designated random (i.e., choosing actions uniformly), while all other nodes used the LLM-driven policy.

We also tested conditions with optional backstory cues in the agent prompts (e.g., labeling an agent as a student or specifying a gender, or both); details of these variants are omitted here.  Full prompt templates, backstory variants that define the persona, and implementation details, including parsing routines, are provided in the Supplementary Information, Section~S1.

\subsection{LLM models and framework}
We evaluate nine open-weight large language models spanning multiple model families, parameter scales, and training paradigms, including variants of LLaMA~\cite{touvron2023llama}, DeepSeek~\cite{bi2024deepseek}, Qwen~\cite{yang2025qwen3}, and SmolLM~\cite{allal2025smollm2}. All models were selected to represent a broad spectrum of open-source LLMs commonly used in agent-based simulations. The evaluated models belong to four open-weight families. LLaMA models are decoder-only transformer language models developed by Meta, available in multiple generations and sizes. In this study, we use both standard instruction-tuned variants (i.e., llama3.1:8b, llama3.1:70b, and llama4:16x17b) and alignment-modified derivatives (i.e., BlackHillsInfoSec llama-3.1:8b-abliterated and meta-llama-3.1:70b-instruct-abliterated IQ4 XS), allowing us to assess whether differences in model scaling and instruction tuning affect emergent cooperative behavior in multi-agent simulations. In particular, we compare standard instruction-tuned LLaMA models with modified ``abliterated'' variants of the same base architecture, which serve as alignment-perturbed counterparts to evaluate the sensitivity of the results to behavioral steering. SmolLM models are compact open-weight language models designed for efficient inference and deployment. Here we test smolLM2.1:7b. DeepSeek models are open-weight language models released in multiple parameter scales and variants, and we tested two different sizes: deepseek-llm:7b and deepseek-llm:67b. Qwen models are multilingual instruction-tuned large language models, with recent variants (e.g., Qwen3) showing strong performance on reasoning and general language understanding benchmarks, here we use qwen3:32b.

\subsection{Quantitative analysis}
We evaluate whether the simulations reproduce the empirical experiments at two complementary levels: (i)\emph{macro} dynamics (i.e., the time evolution of the fraction of cooperators in the network), and (ii) \emph{micro} behavior (i.e., heterogeneity across individuals and context-dependent decision rules). Full definitions, robustness checks, and statistical testing procedures are reported in the Supplementary Information (Section~S2).

\subsubsection{Macro-level agreement}
For each condition and network, we summarize behavior at iteration $t$ by the cooperation rate
\begin{equation}
X^{\mathcal D}(t)=\frac{1}{N}\sum_{i=1}^{N}\mathbf{1}\left[a_i^{\mathcal D}(t)=C\right],
\end{equation}
where $\mathcal D\in\{E,S\}$ denotes empirical ($E$) or simulated ($S$) data, $X^{\mathcal D}(t)$ is the fraction of cooperative agents, and $C$ means cooperation. We then compare empirical and simulated series using two complementary notions of similarity. First, we compute pointwise trajectory errors (RMSE and MAE). These measures quantify how close simulations are in \emph{level} at each iteration, in which smaller values indicate closer agreement, with RMSE penalizing large deviations more. Second, we compute Pearson correlation between the two trajectories, both on the raw series ($r$) and on first differences ($r^{\Delta}$). Correlation captures whether simulations reproduce the \emph{temporal pattern}, where co-movement of increases and decreases, even when small level shifts persist.

\subsubsection{Macro-level temporal dependence}
Trajectory similarity in level or correlation does not guarantee that the simulated process evolves over time in the same way as the empirical one. To compare temporal dependence, we fit a first-order autoregressive model (AR(1))~\cite{hamilton1994time} to the first-differenced series $\Delta X^{\mathcal D}(t)$ (stationarity check~\cite{dickey1979distribution} and the differenced specification are detailed in the Supplementary Information, Section~S2.2.4). For a series $\Delta X^{\mathcal{D}}_{c,g,r}(t)$, the AR(1) model is given by
\begin{equation}
\Delta X^{\mathcal{D}}_{c,g,r}(t) = \kappa + \phi\, \Delta X^{\mathcal{D}}_{c,g,r}(t-1) + \varepsilon(t),
\end{equation}
where $\kappa$ is a constant, $\phi$ captures persistence, and $\varepsilon(t)$ is a white-noise error term. We estimate the model separately for the empirical series $X^{E}_{c,g,r}(t)$ and the simulated series $X^{S}_{c,g,r}(t)$. The coefficient $\phi$ summarizes how strongly changes in cooperation propagate from one iteration to the next: values closer to zero indicate rapid adjustment, whereas larger magnitudes indicate more persistent (or anti-persistent) dynamics. We quantify dynamic similarity using the absolute difference $|\phi_S-\phi_E|$, and we additionally compare the implied mean increment of the differenced process, which captures systematic drift in cooperation over time.

\subsubsection{Significance via null models}
Because the absolute magnitudes of these metrics depend on the scale and length of the series, we assess whether the observed agreement exceeds what would be expected by chance. We construct Monte Carlo null time series that preserve the empirical marginal distribution of cooperation rates while destroying temporal dependence, and we compute two-tailed $p$-values by comparing the observed metric to its null distribution (see Supplementary Information, Section~S2.2.5). For correlations, we test against the standard null hypothesis of zero correlation.

\subsubsection{Micro-level mechanisms}
We quantify individual behavioral heterogeneity by computing each agent's cooperation propensity over the full time horizon $T$. For dataset $\mathcal{D}$, this is defined as
\begin{equation}
    P^{\mathcal{D}}_{i} = \frac{1}{T}\sum_{t=1}^{T}y_i^\mathcal{D}(t),
\end{equation}
where $y_i^\mathcal{D}(t) = \mathbf{1}\!\left[a_i^{\mathcal{D}}(t) = C\right] \in \{0,1\}$ indicates whether agent $i$ cooperated at round $t$.
To test whether simulations reproduce individual-level diversity, we compare the empirical and simulated distributions of $\{P_i\}$ using the first-order Wasserstein distance ($W_1$). Lower values indicate closer agreement in behavioral heterogeneity, rather than merely similar mean cooperation levels (see Supplementary Information, Section~S2.3). To probe context-dependent strategies, we analyze conditional cooperation as a function of (i) an agent's own previous action and (ii) the fraction of cooperating neighbors in the previous round (for details, see Supplementary Information, Section~S2.3.4).


\bmhead{Acknowledgements}
H.F.A. acknowledges ARAID for its financial support. A.A. acknowledges support from the grant RYC2021-033226-I funded by MICIU/AEI/10.13039/501100011033 and the European Union NextGenerationEU/PRTR. A.A. and Y.M. were partially supported by the Government of Arag\'on, Spain, and ERDF ``A way of making Europe'' through grant E36-23R (FENOL), and by Grant No. PID2023-149409NB-I00 from Ministerio de Ciencia, Innovaci\'on y Universidades, Agencia Espa\~nola de Investigaci\'on (MICIU/AEI/10.13039/501100011033) and ERDF ``A way of making Europe''.

\bmhead{Author contributions}
H.F.A.: Conceptualization,
Data curation,
Software,
Formal analysis,
Validation,
Investigation,
Visualization,
Methodology,
Writing-original draft, 
Writing-review and editing.
C.G.L.:
Conceptualization,
Data curation,
Investigation,
Methodology,
Writing-review and editing.
A.A.:
Conceptualization,
Investigation,
Methodology,
Writing-review and editing.
Y.M.: 
Conceptualization,
Formal analysis,
Investigation,
Methodology,
Writing-original draft,
Writing-review and editing.

\bmhead{Competing interests}
The authors declare no competing interests.

\clearpage
\newgeometry{a4paper, margin=1in}
\lstset{
  basicstyle=\ttfamily\small,
  breaklines=true,
  breakatwhitespace=true,
  breakindent=0pt, 
  columns=fullflexible
}

\section*{Supplementary information}


\renewcommand{\thefigure}{S\arabic{figure}}
\setcounter{figure}{0}

\renewcommand{\thetable}{S\arabic{table}}
\setcounter{table}{0}

\renewcommand{\thesection}{S\arabic{section}}
\setcounter{section}{0}

\section{Technical details}
\label{sec:technical-details}

\subsection{Overview}
We simulate a repeated multi-agent Prisoner's Dilemma on an input network. Each node is controlled either by an LLM-based agent or by a randomized policy. The CrewAI orchestration layer is used to orchestrate LLM agents, tasks, and responses, as well as an LLM served via an Ollama-compatible endpoint.

In summary, in each round of the dynamics:
\begin{enumerate}
  \item Each agent receives a brief description of the previous round, including neighbors' choices and normalized payoffs.
  \item Each agent replies with a JSON object containing a \texttt{choice} (``GREEN'' or ``BROWN'') and a free-text \emph{reasoning}.
  \item The simulation computes payoffs for every node based on the payoff matrix, logs the answers, and proceeds to the next round.
\end{enumerate}

As for the graph input and experimental setup, we used the same graphs used in the original paper (i.e.,~\cite{gracia2012heterogeneous}). This includes the network structures and the temporal changes of edges for the control cases, as well as numbers of iterations.

\subsection{Agent creation and random agents}
For each node, we create either:
\begin{itemize}
  \item an LLM-driven \texttt{Agent} (CrewAI), or
  \item a random agent (chooses uniformly between GREEN and BROWN).
\end{itemize}

A fraction $\rho$ of nodes are forced to be random. These nodes remain random throughout the dynamics' execution.

\subsection{Payoff computation}
Payoffs are pairwise and defined by the following matrix (per neighbor):
\[
\begin{array}{c|cc}
\text{You / Neighbor} & \text{GREEN} & \text{BROWN} \\ \hline
\text{GREEN} & 7 & 0 \\
\text{BROWN} & 10 & 0
\end{array}
\]
The node's payoff in a round equals the sum over its neighbors of the payoff according to each neighbor's choice. 

Implementation details:
\begin{itemize}
  \item It iterates over neighbors of node \(i\) and sums the correct payoff depending on the node's choice and the neighbor's choice.
  \item For comparison purposes, a \emph{normalized} neighbor payoff is computed when exposing neighbor payoffs to agents. The normalization rescales a neighbor's payoff by the ratio of the focal node's degree to the neighbor's degree.
\end{itemize}

\subsection{LLM orchestration and prompts}
Important fields passed to each \texttt{Agent} instance include:
\begin{itemize}
  \item \texttt{role}: short role description shown to the model.
  \item \texttt{goal}: short task: ``maximize your own ECUs.'' ECU is a monetary unit.
  \item \texttt{backstory}: personalized per agent (degree and optional backstory fields).
  \item \texttt{system template}: a brief system message explaining the experiment.
  \item \texttt{prompt template}: in the implementation set to:
    \begin{lstlisting}
    Answer the task below in valid JSON format: {input}
    \end{lstlisting}
    The placeholder \texttt{\{input\}} is a dynamic variable that is automatically replaced at runtime with the task-specific description passed to the \texttt{Task} object.
  \item The CrewAI \texttt{Task} is created with expected output describing the required JSON format.
\end{itemize}

The main LLM configuration values used by the script are:
\begin{itemize}
  \item \texttt{llm model} = \texttt{``ollama/llama4:16x17b''} (default, but we also tested several LLMs).
\end{itemize}

\subsubsection{Prompts and textual templates}
The textual templates used are reproduced below. These are sent to the agent together with the task description.

\subsubsection*{Role and goal}
\begin{lstlisting}
role = "You are participating in a social experiment as a volunteer. Your objective is to maximize your total ECUs, the monetary unit. An ECU is a monetary unit"

goal = "Objective: **Your goal is to maximize your own ECUs (ECU is a monetary unit).**\n"
\end{lstlisting}

\subsubsection*{System prompt}
\begin{lstlisting}
system_template = "You are not expected to behave in any particular way. Whatever you do will determine the amount of money you can earn. You have a written version of this direction, which you can check at any stage of the experiment."
\end{lstlisting}

\subsubsection*{Prompt template}
\begin{lstlisting}
"Answer the task below in valid JSON format: {input}"
\end{lstlisting}

\subsubsection*{Expected response format}
All agents must reply with a JSON object in the following format (this is enforced using the \texttt{expected\_output} string in the code):
\begin{lstlisting}
String in a Json format ONLY with choice (GREEN or BROWN) and reasoning (plain text):
{
  "choice": ...,
  "reasoning": "..."
}
\end{lstlisting}

\subsubsection*{First-iteration description (displayed to the agent at round 0)}
A human-readable explanation of payoffs and examples is provided. 
\begin{lstlisting}
Possible payoffs per neighbor.
In the following table each row corresponds to the decision you can make and each column correspond to one of your neighbors' decision.
Your choice / Neighbor choice \t GREEN \t BROWN
GREEN \t 7 \t 0
BROWN \t 10 \t 0

Consider that: you and each of your neighbors (which are humans) will globally earn more if you both choose GREEN (7 ECUs you/7 ECUs your neighbor); you will earn more if you choose BROWN and your neighbor chooses GREEN (10 ECUs you/0 ECUs your neighbor); but if both you and your neighbor choose BROWN you both will earn less (0 ECUs you/0 ECUs your neighbor) than if you both chose GREEN.

This is the screen you will be seeing during the experiment (note that each participant actually sees the graph corresponding to his/her connectivity).
Each round you must choose one of them clicking the corresponding button.

These are some examples of what you could earn in a round.
Example 1: Imagine you choose GREEN, 3 of your neighbors choose GREEN and 1 chooses BROWN. In that round you will earn 3 x 7 + 1 x 0 = 21 ECUs.
Example 2: In another round you choose BROWN, 2 of your neighbors choose GREEN and 2 choose BROWN. In that round you will earn 2 x 10 + 2 x 0 = 20 ECUs.

Round iteration.
Remember that each part will consist of an undetermined number of rounds.
Each round you will have up to 20 s to choose a color. After these 20 s, if you didn't choose, the system will choose for you.
Whatever happens it will not affect the behavior of the system in the next rounds: you will be able to make your subsequent choices normally. (Don't worry: 20 s are more than enough to make a choice).
The round will not end until all participants have made their choice.
At the end of each round you will see a screen like this one. Your choice (as given by the color) and your earning in this round. Also your [NUMBER_OF_NEIGHBORS] neighbors' choices (represented by their colors) and their respective earnings in that round.
Your neighbors' earnings are given with respect to your number of neighbors. For example, you have 5 neighbors and one is Ferdinand (fictitious name). Ferdinand in turn has two neighbors: one is you and the other a stranger. If Ferdinand has won 10 ECUs in the last round, the gain of Ferdinand that you are shown is: (10 ECUs/2 neighbors of Ferdinand) x 5 neighbors of you = 25 ECUs.
Note that what each of your neighbors has won depends on what you have chosen and also on what the neighbors of your neighbors have chosen.
Immediately after finishing a round there will be a new one, and then another one, and so on until you see a screen warning you about the end of that part of the experiment.
\end{lstlisting}

Note that the prompt placeholder \texttt{[NUMBER\_OF\_NEIGHBORS]} is replaced by the number of neighbors. In addition, the prompt explicitly stated that these neighbors corresponded to human participants (``which are humans''), thereby indicating to the LLM that it was interacting with people in a real experiment. This framing substantially affected the resulting behavior.

\subsubsection*{Backstory variants}
The code supports three \texttt{backstory type} values:
\begin{itemize}
  \item \texttt{STUDENT}: adds ``You are a high school student.''
  \item \texttt{GENDER}: adds a ``Gender: man/woman.'' line (based on vertex property \texttt{is\_man} from the data of~\cite{gracia2012heterogeneous}).
  \item \texttt{STUDENT\_AND\_GENDER}: combines the above.
\end{itemize}

\subsubsection*{Round-to-round description}
For rounds $r \ge 1$, the agent receives a per-neighbor summary. The description contains:
\begin{itemize}
  \item The round index.
  \item Each neighbor's previous choice and the neighbor's \emph{normalized} payoff.
  \item The focal node's previous choice and payoff.
  \item The instruction: ``Based on this information, make your choice for this round.''
\end{itemize}

\begin{lstlisting}
Now, you have to answer again:
This is round [ROUND_NUMBER].
In the previous round, your neighbors made the following choices:
Neighbor choices in the previous round:
Neighbor [NEIGHBOR_ID_1]: choice = [NEIGHBOR_CHOICE_1], payoff = [NORMALIZED_NEIGHBOR_PAYOFF_1].
Neighbor [NEIGHBOR_ID_2]: choice = [NEIGHBOR_CHOICE_2], payoff = [NORMALIZED_NEIGHBOR_PAYOFF_2].
Neighbor [NEIGHBOR_ID_3]: choice = [NEIGHBOR_CHOICE_3], payoff = [NORMALIZED_NEIGHBOR_PAYOFF_3].
Your choice and payoff in the previous round was: choice = [YOUR_PREVIOUS_CHOICE], payoff = [YOUR_PREVIOUS_PAYOFF].
Based on this information, make your choice for this round.
\end{lstlisting}

Notice that in the simulations, all prompt placeholders were replaced by their corresponding values. The prompt shown here illustrates an agent interacting with three neighbors.

\subsection{Parsing and error handling}
\begin{itemize}
  \item The script expects the agent response to be valid JSON. Because LLMs occasionally output non-strict JSON, the code contains helpers to sanitize and robustly parse responses:
    \begin{itemize}
      \item \emph{Normalize json string(s)}: trims text outside the outermost \{...\}, removes control characters, attempts to fix common issues (missing braces, extra whitespace between braces, etc.), and finally ensures the string looks like a JSON object.
      \item \emph{Convert string to json}: wraps \texttt{json} loads and, when necessary, raises a descriptive error on failure.
      \item \emph{String to choice output (json string)}: uses the normalizer, parses the JSON, and returns a Python dict with keys \texttt{choice} and \texttt{reasoning}.
    \end{itemize}
    Note that this is not the standard implementation. However, we implemented it this way to avoid minor issues common to less sophisticated LLMs.
  \item If parsing or the Crew kickoff call fails, the code retries up to \texttt{max\_tries} times (default \texttt{max\_tries = 10}). On repeated failure, the implementation falls back to:
    \begin{itemize}
      \item a default response \texttt{\{"choice": "GREEN", "reasoning": "Error: ... defaulting to GREEN."\}}; or
    \end{itemize}
\end{itemize}

This error message was implemented as a precaution to ensure robustness across different LLMs. In practice, however, such parsing failures were not observed for \texttt{llama4\_16x17b}, the model used in the main experiments.

\subsection{Simulation: main loop and execution logic}

The simulation proceeds for \texttt{round\_num = 0, \dots, max\_rounds-1}.
Each round consists of the following steps:

\begin{enumerate}
    \item \textbf{Description preparation.}
    \begin{itemize}
        \item For round 1, agents receive a full explanation of the payoff matrix,
        examples, and the experimental structure.
        \item For subsequent rounds ($r > 1$), agents receive:
        \begin{itemize}
            \item each neighbor's previous choice,
            \item each neighbor's normalized payoff,
            \item their own previous choice and payoff.
        \end{itemize}
    \end{itemize}

    \item \textbf{Task construction.}  
    A \texttt{Task} object is created for each non-random agent. The task
    description is injected into the prompt template through the
    \texttt{\{input\}} placeholder, ensuring that the round-specific information
    is embedded into a fixed JSON-format instruction.

    \item \textbf{Decision generation.}  
    For each node $i$:

    \begin{itemize}
        \item If the node is random, a uniform random choice between GREEN
        and BROWN is generated.
        \item Otherwise, the CrewAI pipeline is executed:
        \begin{lstlisting}
response = Crew(...).kickoff()
parsed = string_to_choice_output(response.raw)
        \end{lstlisting}
    \end{itemize}

    \item \textbf{Retry mechanism and robustness.}

    LLM calls may fail for two reasons:
    \begin{enumerate}
        \item \textbf{Server or connection failures} (e.g., temporary
        unavailability of the Ollama endpoint).
        \item \textbf{Formatting errors} (the LLM does not produce valid JSON).
    \end{enumerate}

    To ensure robustness, each agent call is retried up to \texttt{max\_tries = 10} times.

    This retry mechanism is important because:
    \begin{itemize}
        \item Server-side failures occasionally occur in long simulations.
        \item We empirically observed that when formatting errors occur, they are significantly more frequent in the \emph{first round}.
        In later rounds, once the model has produced a correctly formatted JSON response, formatting mistakes become much less frequent.
    \end{itemize}

    If all retries fail, the system assigns a safe default response:
    \begin{lstlisting}
{"choice":"GREEN", "reasoning":"Error... defaulting to GREEN."}
    \end{lstlisting}
    This ensures that the simulation continues without interruption.

    \item \textbf{Payoff computation.}  
    After all nodes have chosen, payoffs are computed using the pairwise payoff matrix.

\end{enumerate}

\subsubsection*{High-level structured pseudocode}

\begin{lstlisting}
initialize graph(s) g
assign exactly floor(rho * N) random agents
create LLM agents for non-random nodes

for round_num in 0 .. max_rounds-1:
    prepare round-specific description

    for each node i:
        if random:
            result = random_choice()
        else:
            retry up to 10 times:
                call LLM
                attempt JSON parsing
            if all retries fail:
                result = default GREEN

        log result

    compute payoffs
    store payoffs in the graph
    update cooperation statistics
end
\end{lstlisting}

\section{Quantitative analysis}
\subsection{Definitions}
\begin{itemize}
    \item Sources of data:
    \begin{itemize}
        \item Empirical: $E$.
        \item Simulations: $S$ (LLM players and LLM $+$ random players).
    \end{itemize}
    \item Notation:
    \begin{itemize}
        \item $\mathcal{D} \in \{E,S\}$: dataset kind
        \item $c$: Experimental condition (student, gender, and student $+$ gender).
        \item $g$: Network structure.
        \item $N$: Number of network nodes.
        \item $r$: Index for number of trials or replications in simulation.
        \item $i$: Agent index.
        \item $t$: Time step.
        \item $a_i^\mathcal{D}(t)$: action $\{C,D\}$ at time $t$ for dataset $\mathcal{D}$, where $C$ and $D$ are cooperate and defect, respectively.
    \end{itemize}
\end{itemize}

Let's define the indicator variable:
\begin{equation}
    y_i^\mathcal{D}(t) = \mathbf{1}\left[a_i^{\mathcal{D}}(t) = C\right] \in \{0,1\},
\end{equation}
thus, $y=1$ means cooperation.

\subsection{Macro analysis}

\textbf{Cooperation rate (macro):}
For each condition $c$, topology $g$, and trial $r$, compute:
\begin{equation}
    X^{\mathcal{D}}_{c,g,r}(t) = \frac{1}{N}\sum_{i=1}^{N}y_i^\mathcal{D}(t),
\end{equation}
which is the fraction of cooperators at time $t$.

\textbf{Cooperation propensity per agent (micro):}
\begin{equation}
    P^{\mathcal{D}}_{i} = \frac{1}{T}\sum_{t=1}^{T}y_i^\mathcal{D}(t),
\end{equation}
i.e., how often an agent $i$ cooperates over the full temporal horizon $T$.

\subsubsection{Time evolution analysis}
Here, we test whether the model can reproduce the evolution of cooperation over time. To accomplish this, we compare simulated and empirical series. First, we compute the trajectory errors to determine whether they are equal at the same iteration. Then, we examine the temporal dependence structure of the empirical data using an autoregressive model.

\subsubsection{Comparison of trajectory errors}
Before using an AR model to examine the temporal dynamics, we assess how closely the simulated series replicates the empirical series in terms of point-by-point accuracy. Two common metrics are used:

\begin{itemize}
    \item \textbf{Root Mean Squared Error (RMSE)}:
    \begin{equation}
        \text{RMSE} = \sqrt{\frac{1}{T} \sum_{t=1}^{T} \left(X^{E}_{c,g,r}(t) - X^{S}_{c,g,r}(t) \right)^2},
    \end{equation}
    which penalizes larger deviations more heavily.
    
    \item \textbf{Mean Absolute Error (MAE)}:
    \begin{equation}
        \text{MAE} = \frac{1}{T} \sum_{t=1}^{T} \left| X^{E}_{c,g,r}(t) - X^{S}_{c,g,r}(t) \right|,
    \end{equation}
    which measures the average absolute deviation and is less sensitive to outliers.
\end{itemize}

While RMSE and MAE provide a measure of trajectory similarity between the two series, they do not capture the underlying temporal dependence structure. Therefore, we complement this analysis with AR model estimation to assess the persistence properties of the series.

\subsubsection{Correlation analysis}
In addition to trajectory errors, we also compute the Pearson correlation coefficient between the empirical and simulated time series. While RMSE and MAE measure the magnitude of deviations at each time step, they do not capture whether the simulated series follows the same temporal pattern as the empirical data. Pearson correlation quantifies the degree of linear association between the two series, indicating whether increases and decreases in cooperation occur synchronously over time. We calculate Pearson both for the raw correlation series ($r$) and for the first differences of the series ($r^{\Delta}$) to capture correlations in the changes between consecutive time steps. For each correlation, we report the corresponding $p$-value to evaluate the statistical significance of the observed association, providing a measure of confidence that the observed correlation is not due to random chance.

\subsubsection{Autoregressive comparison between empirical and simulated series}
Before estimating the autoregressive model, we verify whether the empirical and simulated time series are stationary. Stationarity ensures that the mean, variance, and autocovariance structure of the series do not change over time, which is a fundamental assumption for AR model estimation. To do this, we apply the Augmented Dickey-Fuller (ADF) test to both series~\cite{dickey1979distribution}. 
In our analysis, we estimate the model using the first-differenced series, $\Delta X^{\mathcal{D}}_{c,g,r}(t)$, rather than the level series $X^{\mathcal{D}}_{c,g,r}(t)$. The differenced specification captures changes in cooperation over time and satisfies the stationarity requirement, whereas the level series exhibits non-stationary behavior according to the ADF test.

To assess whether the simulated time series reproduces the temporal dependence structure of the empirical data, we estimate an autoregressive model of order one (AR(1)) for the differenced series~\cite{hamilton1994time}. For a series $\Delta X^{\mathcal{D}}_{c,g,r}(t)$, the AR(1) model is given by
\begin{equation}
\Delta X^{\mathcal{D}}_{c,g,r}(t) = \kappa + \phi\, \Delta X^{\mathcal{D}}_{c,g,r}(t-1) + \varepsilon(t),
\end{equation}
where $\kappa $ is a constant, $\phi$ measures persistence, and
$\varepsilon(t)$ is a white noise error term. We estimate the model separately for the empirical differenced series $\Delta X^{E}_{c,g,r}(t)$ and the simulated series $\Delta X^{S}_{c,g,r}(t)$. The key parameter of interest is the persistence coefficient $\phi$. To evaluate whether the simulation reproduces the dynamic structure, we use the following analysis metrics:

\begin{itemize}
\item \textbf{Difference in persistence coefficients:}
To assess whether the simulated series reproduces the temporal dependence structure of the empirical data, we compare the estimated AR(1) persistence parameters through the absolute difference
\begin{equation}
\text{Diff}_{\phi} = |\phi_{S} - \phi_{E}|.
\end{equation}
The coefficient $\phi$ measures the strength and direction of temporal dependence. If $|\phi_{S} - \phi_{E}|$ is small, the simulation reproduces the degree of persistence (or anti-persistence) observed in the empirical data. Large differences indicate that the simulated dynamics adjust either too quickly or too slowly relative to the empirical process, even if the overall trajectories appear similar.

\item \textbf{Implied mean increment:}
For a stationary AR(1) process in first differences ($|\phi_{\mathcal{D}}|<1$), the unconditional mean of the differenced process is
\begin{equation}
\mu_{\mathcal{D}} = \frac{\kappa_{\mathcal{D}}}{1 - \phi_{\mathcal{D}}}.
\end{equation}
This quantity represents the average change around which the series increments fluctuate over time. Comparing $\mu_E$ and $\mu_S$ allows us to verify whether simulated and empirical processes exhibit similar average dynamics. Substantial differences indicate systematic biases in the evolution of cooperation, even if persistence properties
are similar.
\end{itemize}

\subsubsection{Null model}
Since the magnitude of the proposed metrics cannot be directly interpreted, we use a null model to evaluate their relevance. The null model generates a synthetic time series of the same length as the empirical and simulated series. Observations are independently drawn from the empirical marginal distribution. This procedure preserves the unconditional distribution of cooperation levels while eliminating any temporal dependence structure.

Let $M^{obs}$ denote the observed value of a metric. To construct a reference distribution under the null hypothesis of no temporal structure, we generate $B$ independent null realizations and compute the corresponding metric values $\{M^{(1)}, \dots, M^{(B)}\}$. Next, the statistical significance is assessed by comparing $M^{obs}$ to the empirical distribution of $\{ M^{(b)} \}_{b=1}^{B}$. Here, we use $B=10,000$. The null hypothesis is rejected when $M^{obs}$ lies in the tails of the null distribution.

The Monte Carlo two-tailed $p$-value is defined as
\begin{equation}
p = \frac{1}{B} \sum_{b=1}^{B} \mathbf{1}\Big( \big| M^{(b)} - \mu_{\text{null}} \big| \ge \big| M^{obs} - \mu_{\text{null}} \big| \Big),
\label{eq:pvalue}
\end{equation}
where $\mu_{\text{null}} = \frac{1}{B} \sum_{b=1}^{B} M^{(b)}$ is the mean of the null distribution. Here, we adopt this approach because a small $p_{\text{two-tailed}}$ indicates that the observed metric is unlikely under the null hypothesis, given deviations in either direction. Note that for Pearson correlation, we adopted the null hypothesis of zero correlation ($H_0: r = 0$).

We adopt this two-tailed criterion because our goal is to determine whether the observed metric is unusually far from the null expectation, regardless of direction. Under this interpretation, both unusually large and unusually small values of $M^{\mathrm{obs}}$ relative to the null distribution are treated as evidence against the null hypothesis. Thus, a small $p$-value indicates that the observed metric is unlikely under the null model, not only when it exceeds the null expectation, but more generally when it represents an extreme deviation from it.

\subsection{Micro analysis}
\subsubsection{Definitions}
\textbf{Switching rate and persistence (micro):}
Let's define switching rate: 
\begin{equation}
    S^{\mathcal{D}}_{i} = \frac{1}{T-1}\sum_{t=2}^{T}\mathbf{1}\left[a^{\mathcal{D}}_{i}(t) \neq a^{\mathcal{D}}_{i}(t-1)  \right].
\end{equation}
Persistence is defined as $\mathcal{P}^{\mathcal{D}}_{i}=1-S^{\mathcal{D}}_{i}$.

\textbf{Neighborhood cooperation (for conditional cooperation):}
For each agent $i$ and for $t\geq2$, we can calculate the previous round cooperation in the agent's neighborhood $\mathcal{N}$ as
\begin{equation}
m^{\mathcal{D}}_{i}(t-1)
=
\frac{1}{\left|\mathcal{N}_{g_{t-1}}(i)\right|}
\sum_{j \in \mathcal{N}_{g_{t-1}}(i)}
y_j^{\mathcal{D}}(t-1),
\end{equation}
where $\mathcal{N}_{g_{t-1}}(i)$ denotes the set of neighbors of agent $i$
in the interaction graph $g_{t-1}$.
Also for agent $i$, we can use another variable $z_i^\mathcal{D}(t-1) = y_i^\mathcal{D}(t-1)$ so that with $z_i^\mathcal{D}(t-1)$ and $m^{\mathcal{D}}_{i}(t-1)$ we can study conditional cooperation.

For the conditional-rule analysis, the $p$-value is computed using the same Monte Carlo criterion defined in Eq.~\ref{eq:pvalue}. However, each observation is represented by the triplet $\big(m_i^{\mathcal D}(t-1), z_i^{\mathcal D}(t-1), y_i^{\mathcal D}(t)\big)$. Under the null hypothesis that $y_i^{\mathcal D}(t)$ is independent of the previous-round context, null realizations are generated by randomly permuting the values of $y_i^{\mathcal D}(t)$ across observations while keeping $\big(m_i^{\mathcal D}(t-1)$ and $z_i^{\mathcal D}(t-1)\big)$ fixed. This preserves the marginal distribution of current actions while removing any association between $y_i^{\mathcal D}(t)$ and the predictors $\big(m_i^{\mathcal D}(t-1)$ and $z_i^{\mathcal D}(t-1)\big)$.

\subsubsection{Confidence interval}
Due to computational constraints, for most tests, only one simulation per condition is performed. In this case, confidence intervals are estimated via bootstrapping across agents rather than simulation runs, capturing uncertainty due to agent heterogeneity.

For the confidence interval (CI), we can use a bootstrap method to set pointwise uncertainty. In practice:
\begin{itemize}
    \item Sample $N$ agents/players with replacement (i.e., a player can appear multiple times).
    \item Recompute $\theta^{\mathcal{D}}(t)$ using the sampled agents.
    \item Repeat $B$ times to obtain bootstrap replicates $\left\{\theta^{\mathcal D,(b)}_{c,S}(t) \right\}_{b=1}^{B}$.
    \item Define $CI_{95\%}(t)=\left[Q_{0.025}, Q_{0.975}\right]$.
\end{itemize}

To quantitatively measure the difference between these trajectories, we use RMSE and MAE to measure the distance between the trajectories $\overline{\theta}^{E}_{c,S}$ and $\overline{\theta}^{S}_{c,S}$. In addition, we calculate the Pearson correlation coefficient for both for the raw correlation series ($r$) and for the first differences of the series ($r^{\Delta}$) between the trajectories $\overline{\theta}^{E,(b)}_{c,S}$ and $\overline{\theta}^{S,(b)}_{c,S}$ to capture correlations in the changes between consecutive time steps.

For cases where we run the dynamics $R$ times, we use the Cluster bootstrap to account for observations being grouped by simulation run. Specifically, for each bootstrap replicate, we first resample runs with replacement, then resample agents within each selected run. We then recompute the statistic of interest at each timestep. Repeating this many times yields a bootstrap distribution from which we take the percentiles to have the 95\% confidence interval. This procedure preserves both within-run dependence and between-run variability, providing more realistic uncertainty than pooling all agents across runs as independent observations.

\subsubsection{Heterogeneity across individuals}
To evaluate whether simulations reproduce the diversity of behavioral patterns observed in empirical data, we compare the distributions of individual cooperation propensities across agents. Let
\[
\{P_i^{E}\}_{i=1}^{N}
\quad \text{and} \quad
\{P_i^{S}\}_{i=1}^{N}
\]
denote the cooperation propensities computed for empirical and simulated datasets, respectively. These quantities capture the frequency with which each agent cooperates over the full temporal horizon and therefore encode behavioral heterogeneity at the individual level.

We quantify the discrepancy between both distributions using the first-order Wasserstein distance. First, we sort both samples in non-decreasing order:
\[
P_{(1)}^{\mathcal D} \le P_{(2)}^{\mathcal D} \le \dots \le P_{(N)}^{\mathcal D},
\qquad \mathcal D \in \{E,S\}.
\]
The empirical Wasserstein distance of order one between the two samples
is then defined as
\begin{equation}
W_1(E,S)
=
\frac{1}{N}
\sum_{k=1}^{N}
\left|
P_{(k)}^{E}
-
P_{(k)}^{S}
\right|.
\end{equation}
This metric measures the minimal average amount of probability mass that must be transported to transform the simulated distribution into the empirical one. A small value of $W_1(E,S)$ indicates that the simulation successfully reproduces the level of behavioral heterogeneity observed among real agents, whereas larger values reveal discrepancies such as overly homogeneous or excessively polarized simulated behaviors.

\subsubsection{Conditional cooperation rule}
The conditional cooperation rule is defined as the probability that agent $i$ cooperates at time $t$ given its previous action and the cooperation level in its neighborhood:
\begin{equation}
\Pr\!\left(
y_i^{\mathcal D}(t)=1
\mid
m_i^{\mathcal D}(t-1),
z_i^{\mathcal D}(t-1)
\right),
\end{equation}
recall that
\[
z_i^{\mathcal D}(t-1)=y_i^{\mathcal D}(t-1).
\]

Let $\{b_k\}_{k=0}^{K}$ denote the bin boundaries. Define
\begin{equation}
I_1=[b_0,b_1], \qquad I_k=(b_{k-1},b_k] \;\; \text{for } k=2,\dots,K.
\end{equation}
Then, for each previous action $z\in\{0,1\}$, define
\begin{equation}
S_{k,z}
=
\{(i,t):
m_i^{\mathcal D}(t-1)\in I_k,
\;
z_i^{\mathcal D}(t-1)=z
\}.
\end{equation}

The empirical conditional cooperation probability is estimated as
\begin{equation}
\widehat P_{k,z}^{\mathcal D}
=
\frac{1}{|S_{k,z}|}
\sum_{(i,t)\in S_{k,z}}
y_i^{\mathcal D}(t).
\end{equation}
Thus, for each bin $k$ and previous action
$z\in\{0,1\}$, we obtain four conditional cooperation
probabilities:
\[
\widehat P_{k,0}^{E},\;
\widehat P_{k,1}^{E}
\quad \text{: for experiments},
\]
\[
\widehat P_{k,0}^{S},\;
\widehat P_{k,1}^{S}
\quad \text{: for LLMs},
\]
after previous defection ($z=0$) or cooperation ($z=1$).

Finally, we can do as before and compare these quantities using RMSE and MAE and Pearson correlation.

\newpage
\section{Results}

\subsection{Visual analysis}
First, we show the results for $\rho=0$.

\begin{figure}[h!]
    \centering
    \includegraphics[width=0.95\linewidth]{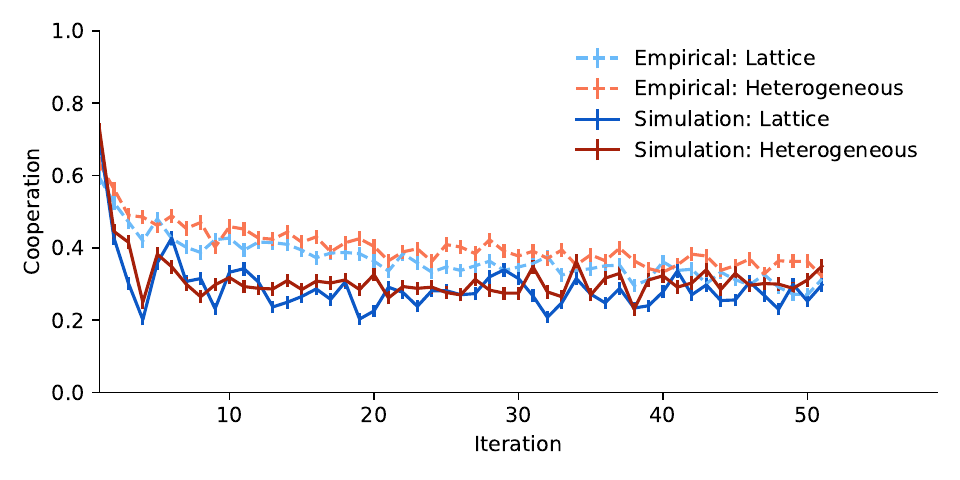}
    \caption{Evolution of cooperation over iterations for the \texttt{STUDENT} condition in the setting. 
    Dashed lines show empirical results from the behavioral experiment, 
    while solid lines show simulation results using \texttt{llama4\_16x17b}. 
    The solid lines represent the mean cooperation rate per round (empirical) or per iteration (simulation), and the error bars represent the standard error. 
    }
    \label{fig:coop_dynamics_student_network}
\end{figure}

\begin{figure}[h!]
    \centering
    \includegraphics[width=0.95\linewidth]{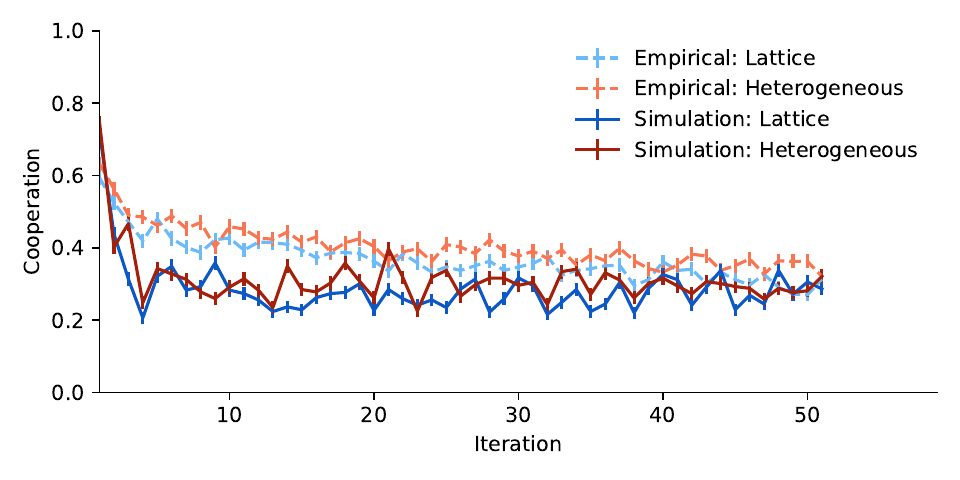}
    \caption{Evolution of cooperation over iterations for the \texttt{GENDER} condition in the setting. 
    Dashed lines show empirical results from the behavioral experiment, 
    while solid lines show simulation results using \texttt{llama4\_16x17b}. 
    The solid lines represent the mean cooperation rate per round (empirical) or per iteration (simulation), and the error bars represent the standard error. 
    }
    \label{fig:coop_dynamics_gender_network}
\end{figure}

\begin{figure}[h!]
    \centering
    \includegraphics[width=0.95\linewidth]{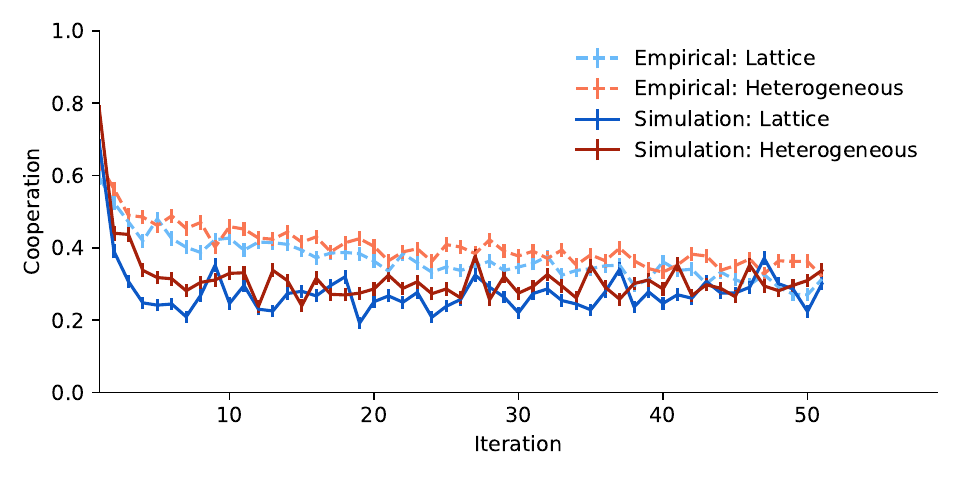}
    \caption{Evolution of cooperation over iterations for the \texttt{STUDENT\_AND\_GENDER} condition  in the setting. 
    Dashed lines show empirical results from the behavioral experiment, while solid lines show simulation results using \texttt{llama4\_16x17b}. The solid lines represent the mean cooperation rate per round (empirical) or per iteration (simulation), and the error bars represent the standard error. 
    }
    \label{fig:coop_dynamics_student_and_gender_network}
\end{figure}

\begin{figure}[h!]
    \centering
    \includegraphics[width=0.95\linewidth]{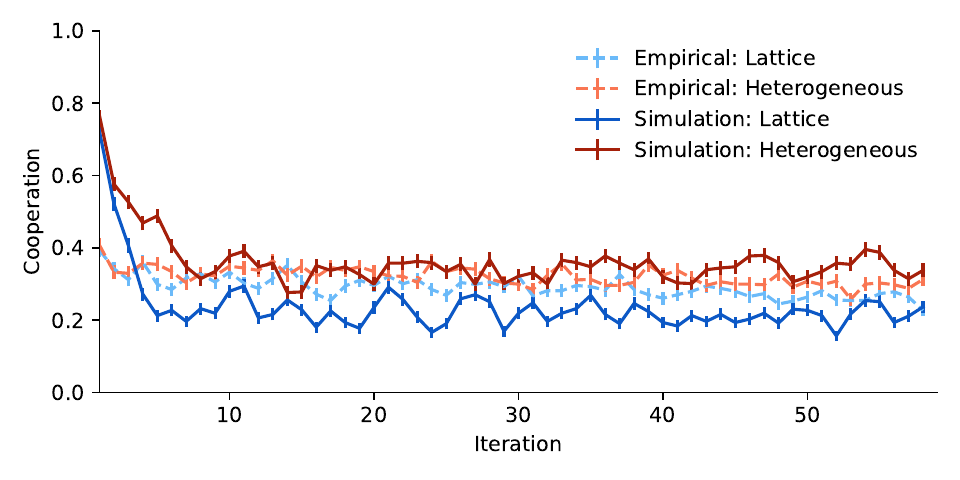}
    \caption{\textbf{Control}: Evolution of cooperation over iterations for the \texttt{STUDENT} condition in the setting. 
    Dashed lines show empirical results from the behavioral experiment, while solid lines show simulation results using \texttt{llama4\_16x17b}. 
    The solid lines represent the mean cooperation rate per round (empirical) or per iteration (simulation), and the error bars represent the standard error. 
    }
    \label{fig:coop_dynamics_student_control}
\end{figure}

\begin{figure}[h!]
    \centering
    \includegraphics[width=0.95\linewidth]{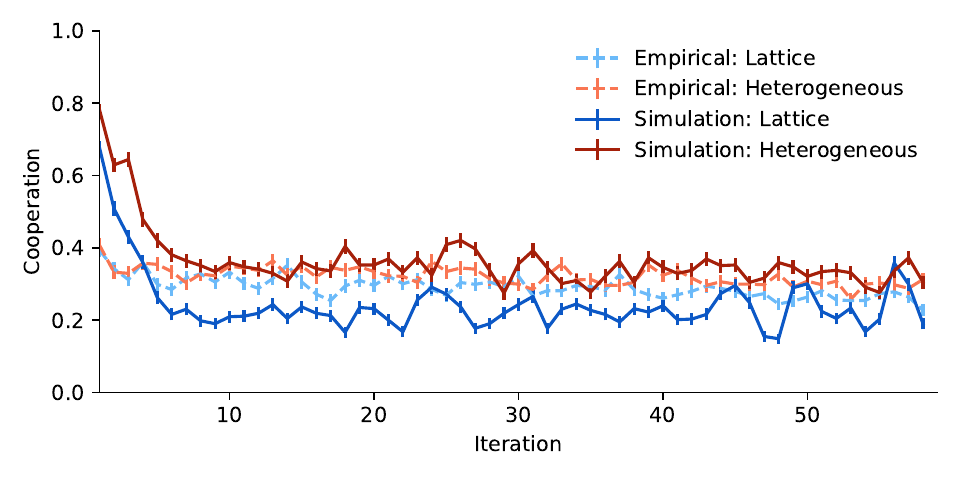}
    \caption{\textbf{Control}: Evolution of cooperation over iterations for the \texttt{GENDER} condition in the setting. 
    Dashed lines show empirical results from the behavioral experiment, 
    while solid lines show simulation results using \texttt{llama4\_16x17b}. 
    The solid lines represent the mean cooperation rate per round (empirical) or per iteration (simulation), and the error bars represent the standard error. 
    }
    \label{fig:coop_dynamics_gender_control}
\end{figure}

\begin{figure}[h!]
    \centering
    \includegraphics[width=0.95\linewidth]{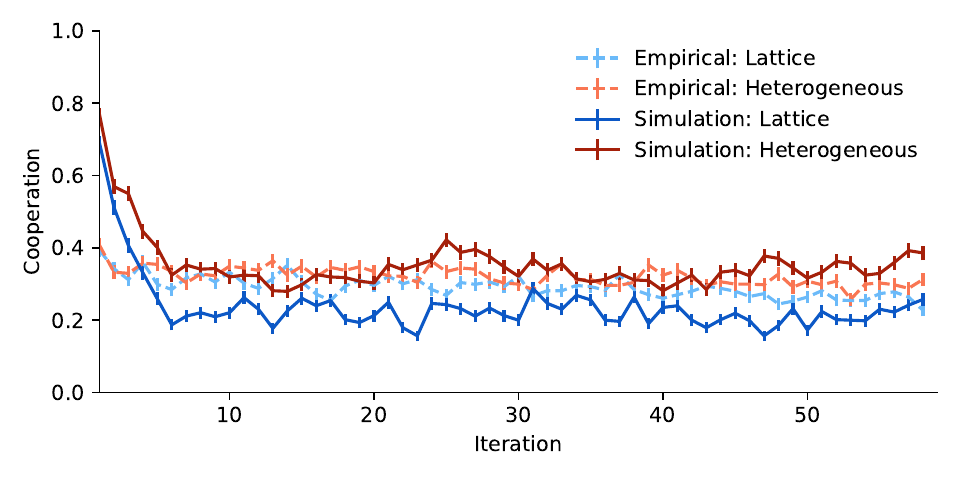}
    \caption{\textbf{Control}: Evolution of cooperation over iterations for the \texttt{STUDENT\_AND\_GENDER} condition  in the setting. 
    Dashed lines show empirical results from the behavioral experiment, 
    while solid lines show simulation results using \texttt{llama4\_16x17b}. 
    The solid lines represent the mean cooperation rate per round (empirical) or per iteration (simulation), and the error bars represent the standard error.  
    }
    \label{fig:coop_dynamics_student_and_gender_control}
\end{figure}

\begin{figure}[h!]
    \centering
    \includegraphics[width=0.95\linewidth]{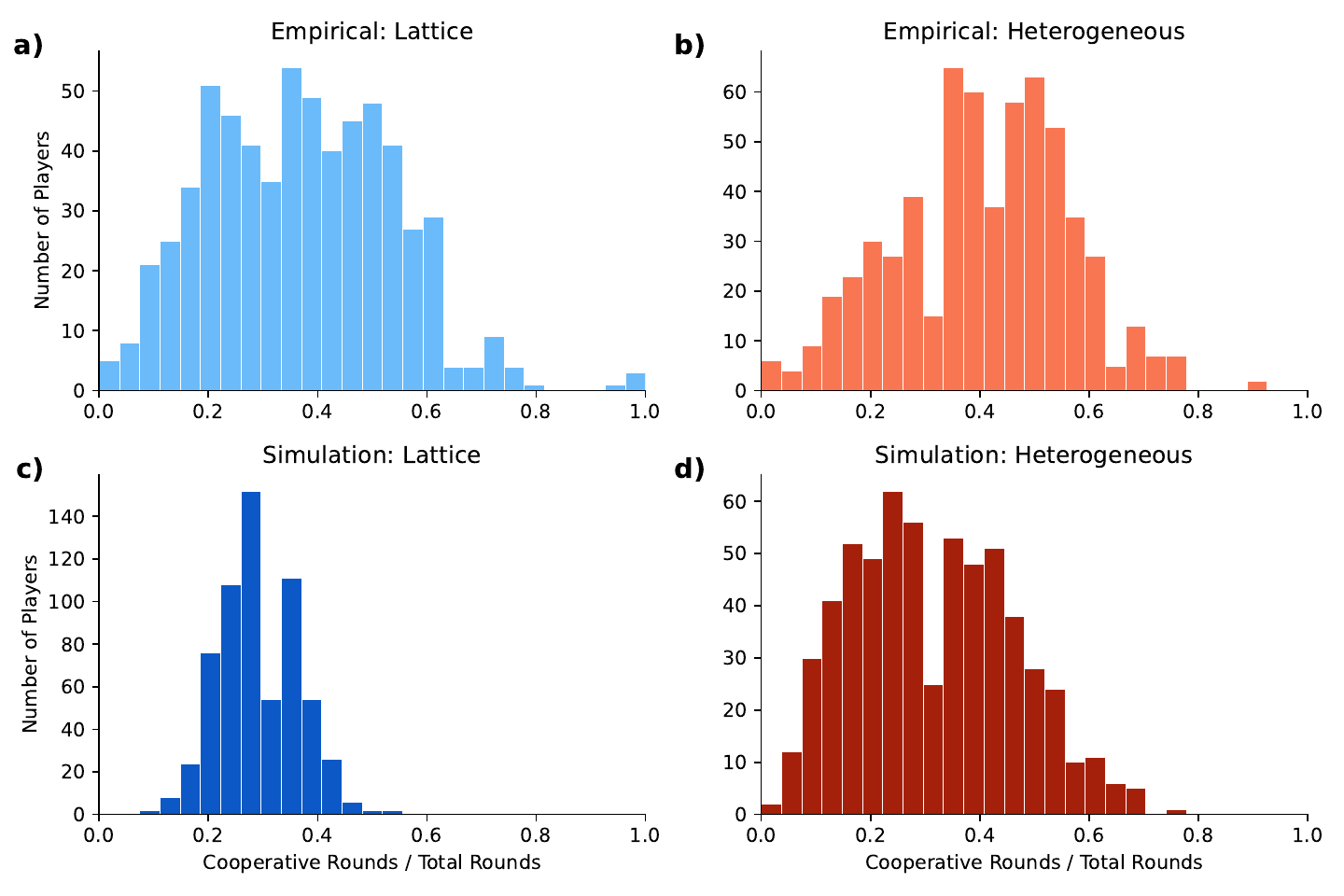}
    \caption{Distribution of per-agent cooperation rates for the \texttt{STUDENT} condition setting. Each histogram displays the fraction of cooperative rounds per player. Bins are uniformly spaced in the interval $[0,1]$. Simulated results are generated using \texttt{llama4\_16x17b}. 
    }
    \label{fig:coop_hist_student_network}
\end{figure}

\begin{figure}[h!]
    \centering
    \includegraphics[width=0.95\linewidth]{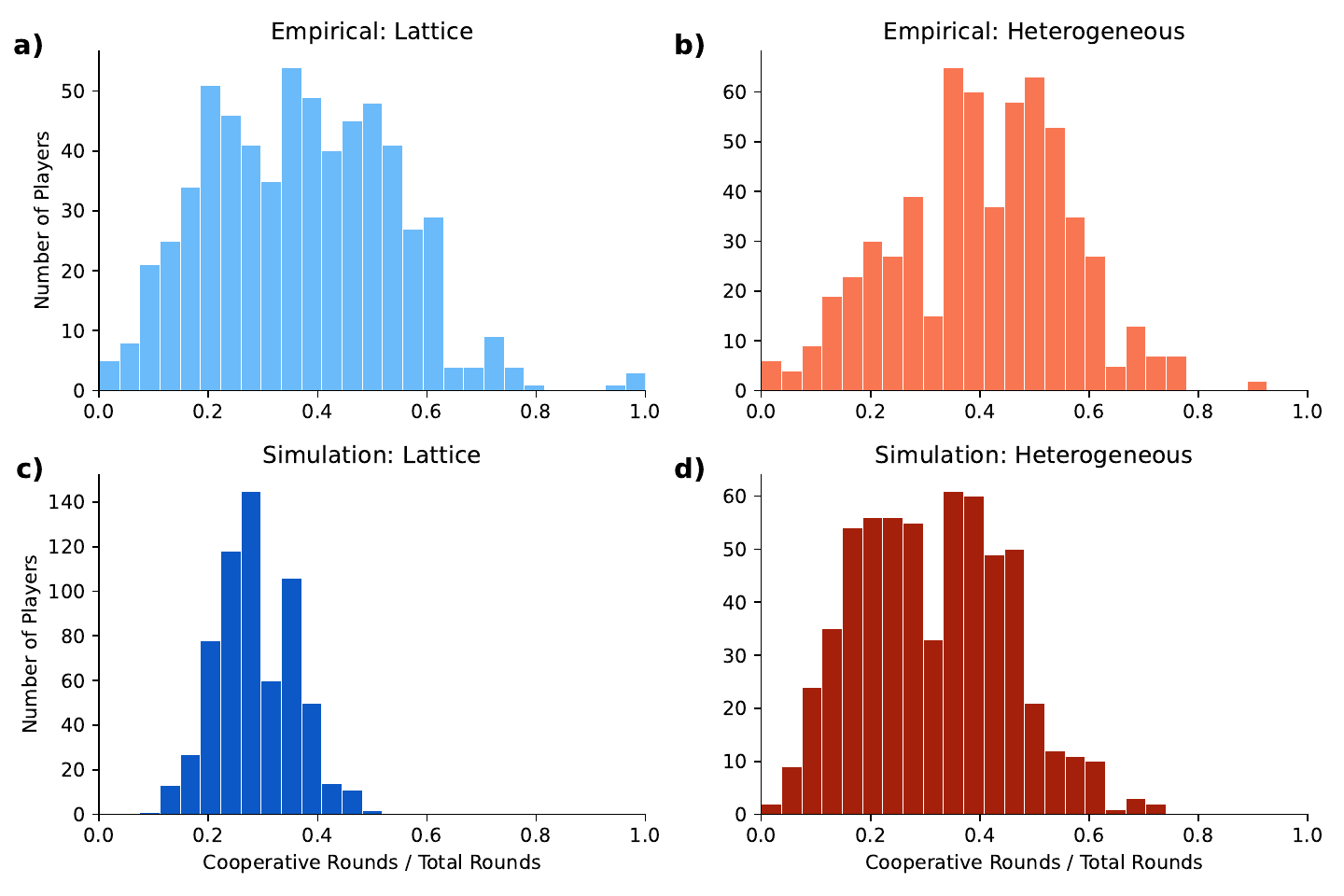}
    \caption{Distribution of per-agent cooperation rates for the \texttt{GENDER} condition setting. Each histogram displays the fraction of cooperative rounds per player. Bins are uniformly spaced in the interval $[0,1]$. Simulated results are generated using \texttt{llama4\_16x17b}. 
    }
    \label{fig:coop_hist_gender_network}
\end{figure}

\begin{figure}[h!]
    \centering
    \includegraphics[width=0.95\linewidth]{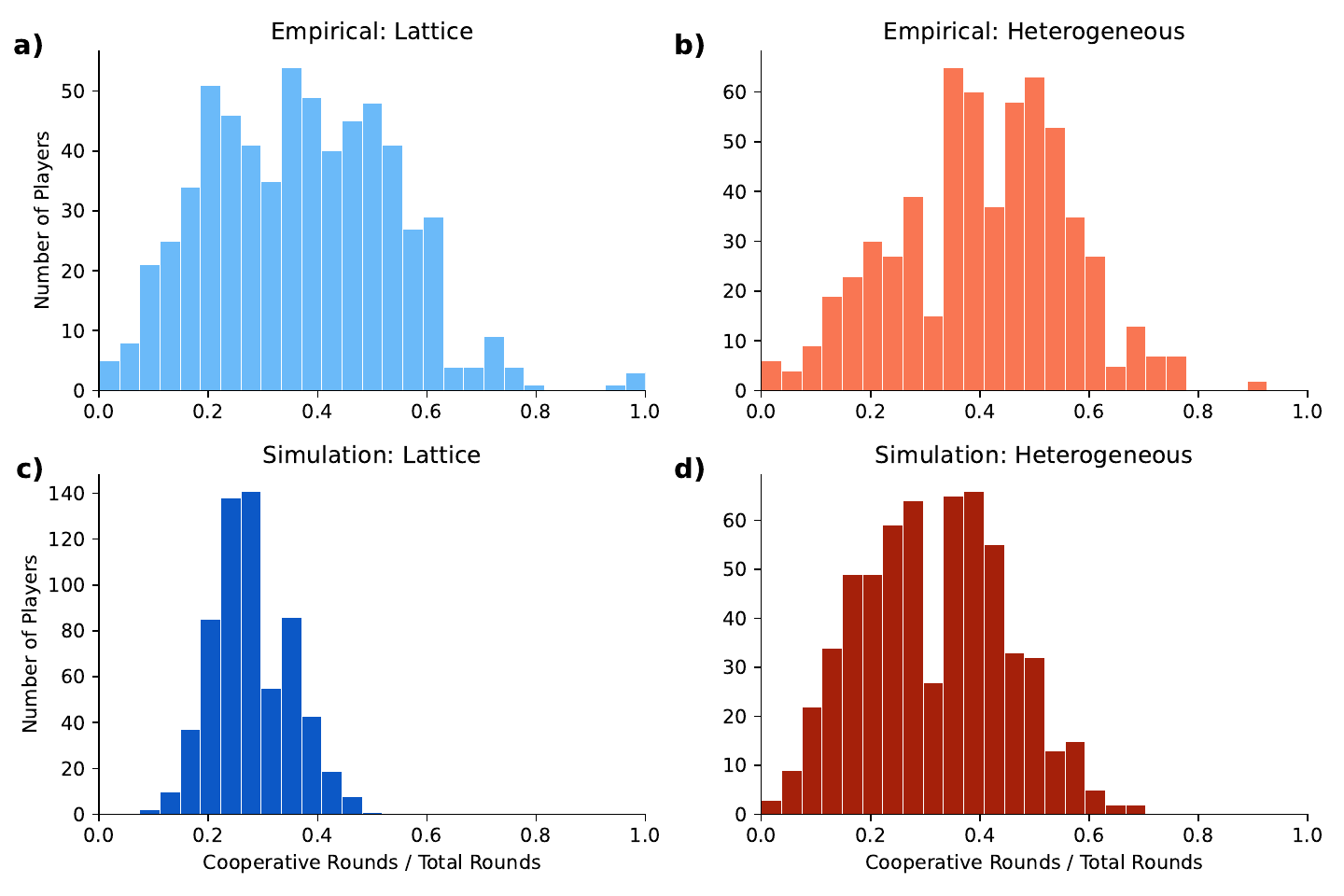}
    \caption{Distribution of per-agent cooperation rates for the \texttt{STUDENT\_AND\_GENDER} condition setting. Each histogram displays the fraction of cooperative rounds per player. Bins are uniformly spaced in the interval $[0,1]$. Simulated results are generated using \texttt{llama4\_16x17b}. 
    }
    \label{fig:coop_hist_student_and_gender_network}
\end{figure}

\begin{figure}[h!]
    \centering
    \includegraphics[width=0.95\linewidth]{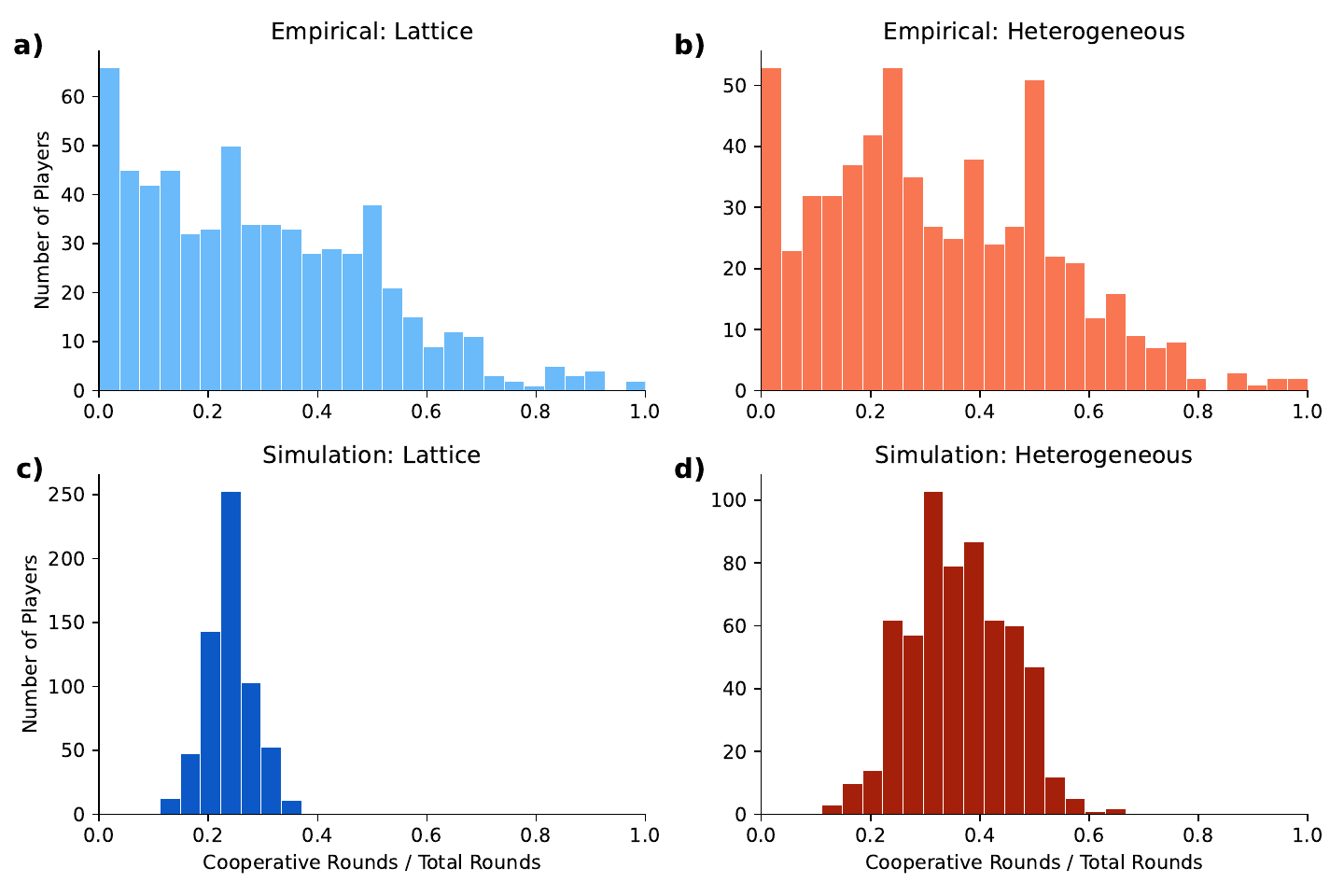}
    \caption{\textbf{Control}: Distribution of per-agent cooperation rates for the \texttt{STUDENT} condition setting. Each histogram displays the fraction of cooperative rounds per player. Bins are uniformly spaced in the interval $[0,1]$. Simulated results are generated using \texttt{llama4\_16x17b}. 
    }
    \label{fig:coop_hist_student_control}
\end{figure}

\begin{figure}[h!]
    \centering
    \includegraphics[width=0.95\linewidth]{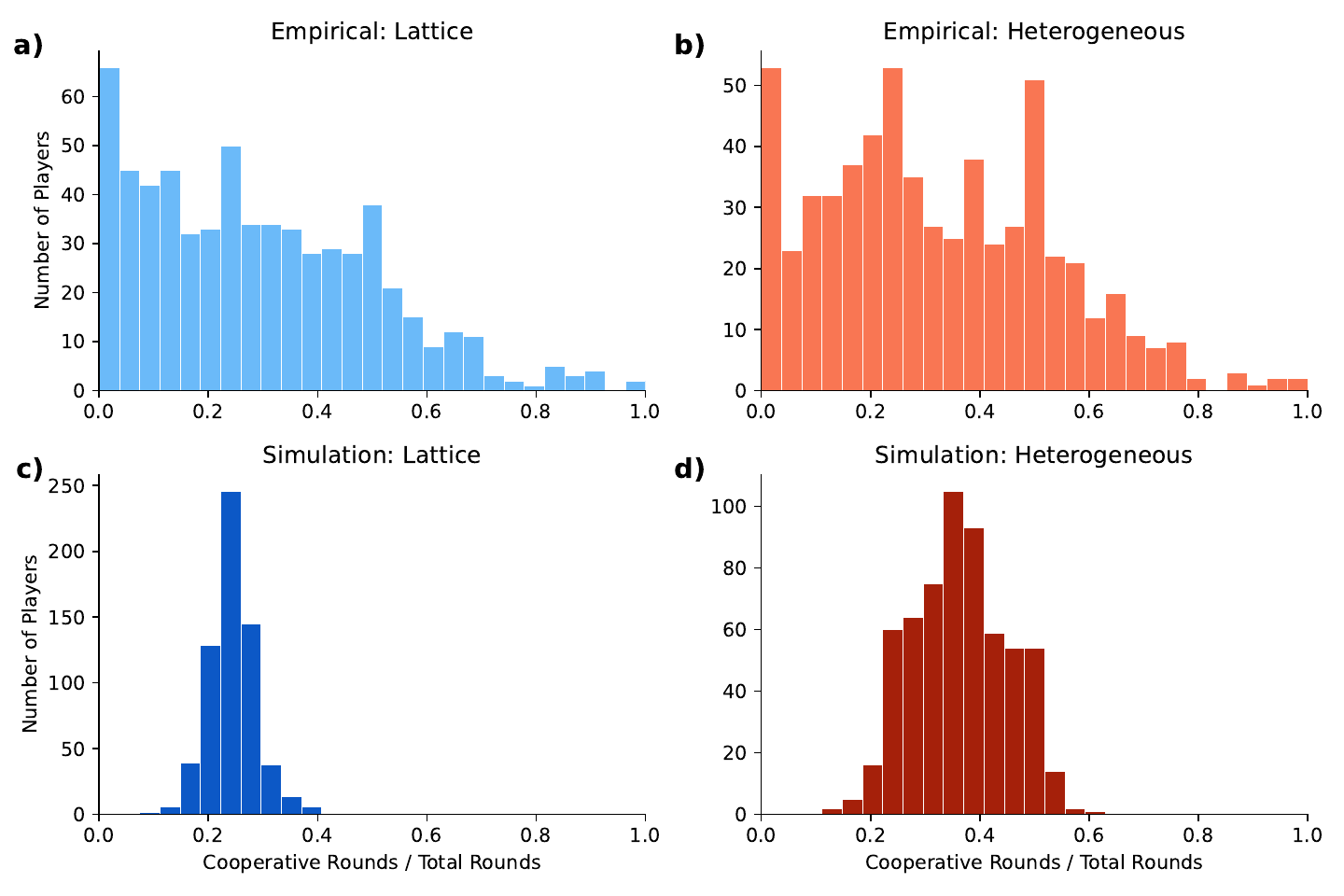}
    \caption{\textbf{Control}: Distribution of per-agent cooperation rates for the \texttt{GENDER} condition setting. Each histogram displays the fraction of cooperative rounds per player. Bins are uniformly spaced in the interval $[0,1]$. Simulated results are generated using \texttt{llama4\_16x17b}. 
    }
    \label{fig:coop_hist_gender_control}
\end{figure}

\begin{figure}[h!]
    \centering
    \includegraphics[width=0.95\linewidth]{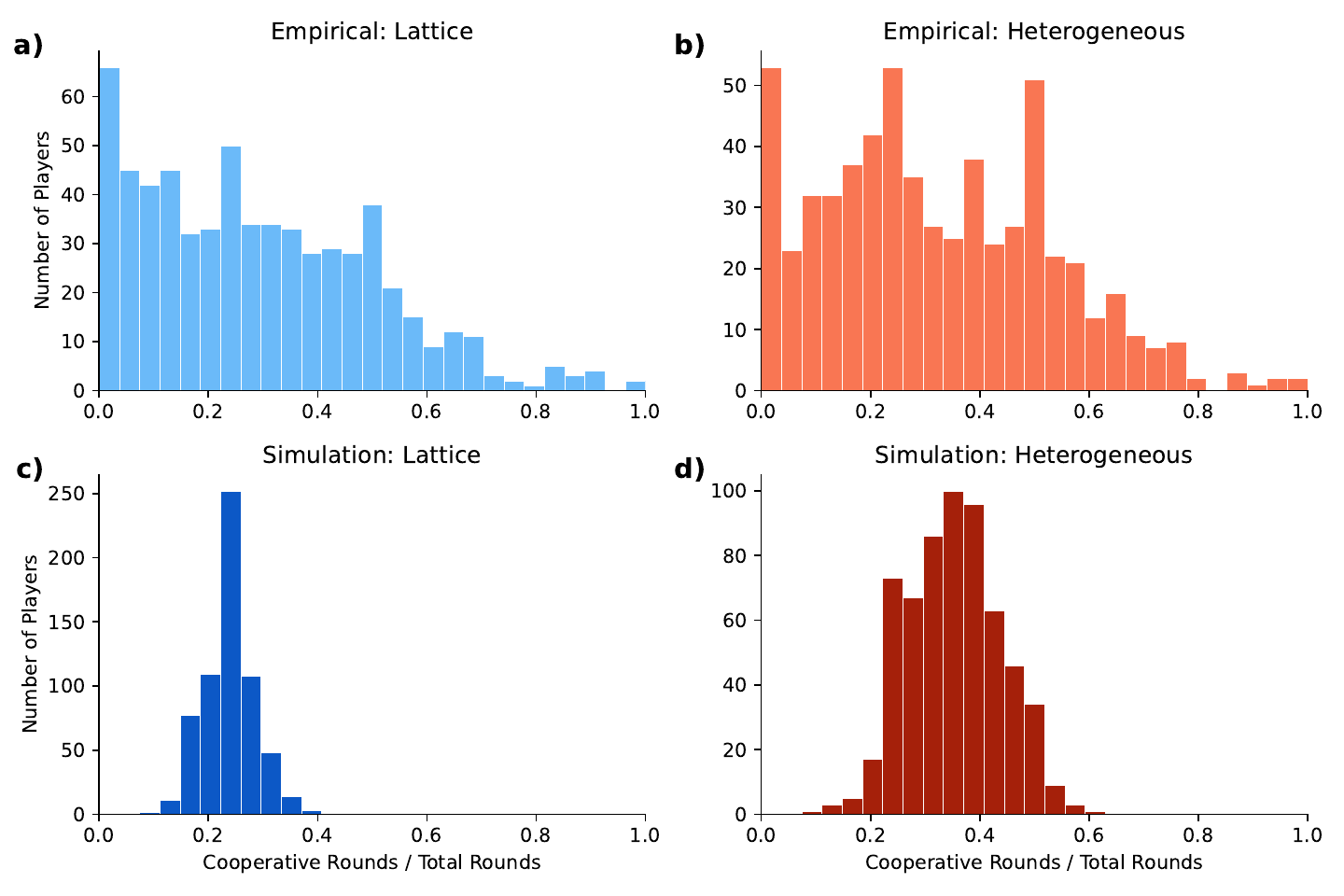}
    \caption{\textbf{Control}: Distribution of per-agent cooperation rates for the \texttt{STUDENT\_AND\_GENDER} condition setting. Each histogram displays the fraction of cooperative rounds per player. Bins are uniformly spaced in the interval $[0,1]$. Simulated results are generated using \texttt{llama4\_16x17b}. 
    }
    \label{fig:coop_hist_student_and_gender_control}
\end{figure}
\FloatBarrier

\begin{figure}[h!]
    \centering
    \includegraphics[width=1.\linewidth]{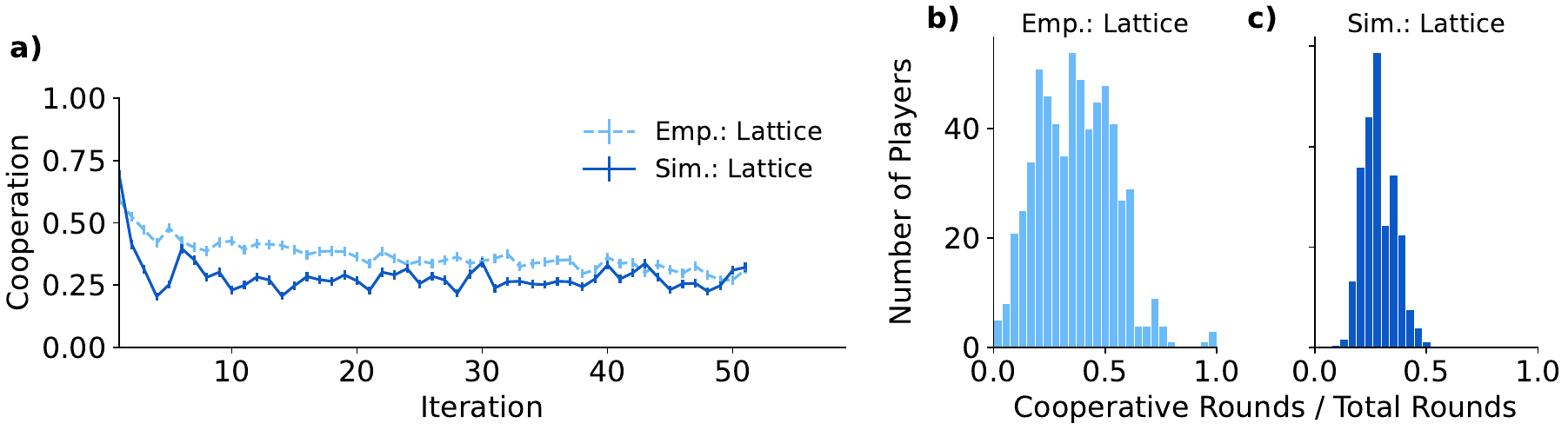}
    \caption{Simulation results obtained with \texttt{llama4\_16x17b} on the Lattice network under different temperatures sampled uniformly from the interval $[0,2]$. 
    }
    \label{fig:temperature}
\end{figure}

In addition to testing different scenarios, we evaluated our simulation setup with \texttt{llama4\_16x17b} across a range of temperatures. Specifically, temperature values were randomly sampled from the interval $[0,2]$. These experiments were conducted exclusively on the lattice network topology (see Figure~\ref{fig:temperature}).

\begin{figure}[h!]
    \centering
    \includegraphics[width=1.\linewidth]{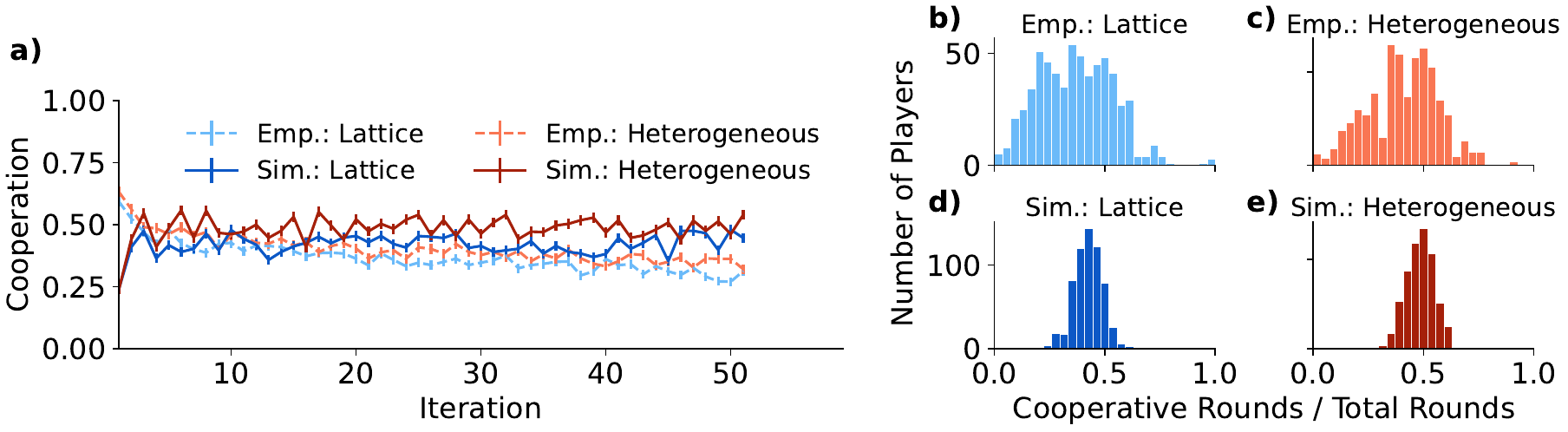}
    \caption{Simulation results obtained with \texttt{qwen3\_32b} on the Lattice and Heterogeneous networks. 
    }
    \label{fig:results_qwen}
\end{figure}

To further assess whether the reduced dispersion observed in the simulations is model-dependent, we repeated the analysis using \texttt{qwen3\_32b}. The results shown in Fig.~\ref{fig:results_qwen}  indicate that the tendency of LLM agents to generate less variable cooperation distributions than human participants is not specific to \texttt{llama4\_16x17b}.

\FloatBarrier

Next, we present the results for other values of $\rho$.

\begin{figure}[h!]
    \centering
    \includegraphics[width=0.95\linewidth]{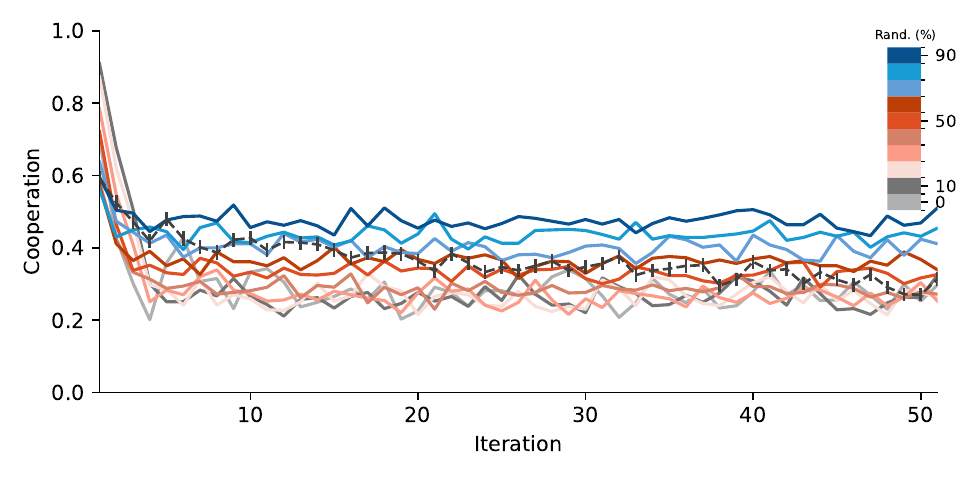}
    \caption{ 
    Cooperation over iterations for the \texttt{lattice} network, varying $\rho$, under the \texttt{STUDENT} condition. The simulation results were generated by \texttt{llama4\_16x17b}. The black dashed line shows the empirical data, and the error bars show the standard error.}
    \label{fig:rand_levels}
\end{figure}
\FloatBarrier

\begin{figure}[h!]
    \centering
    \includegraphics[width=0.95\linewidth]{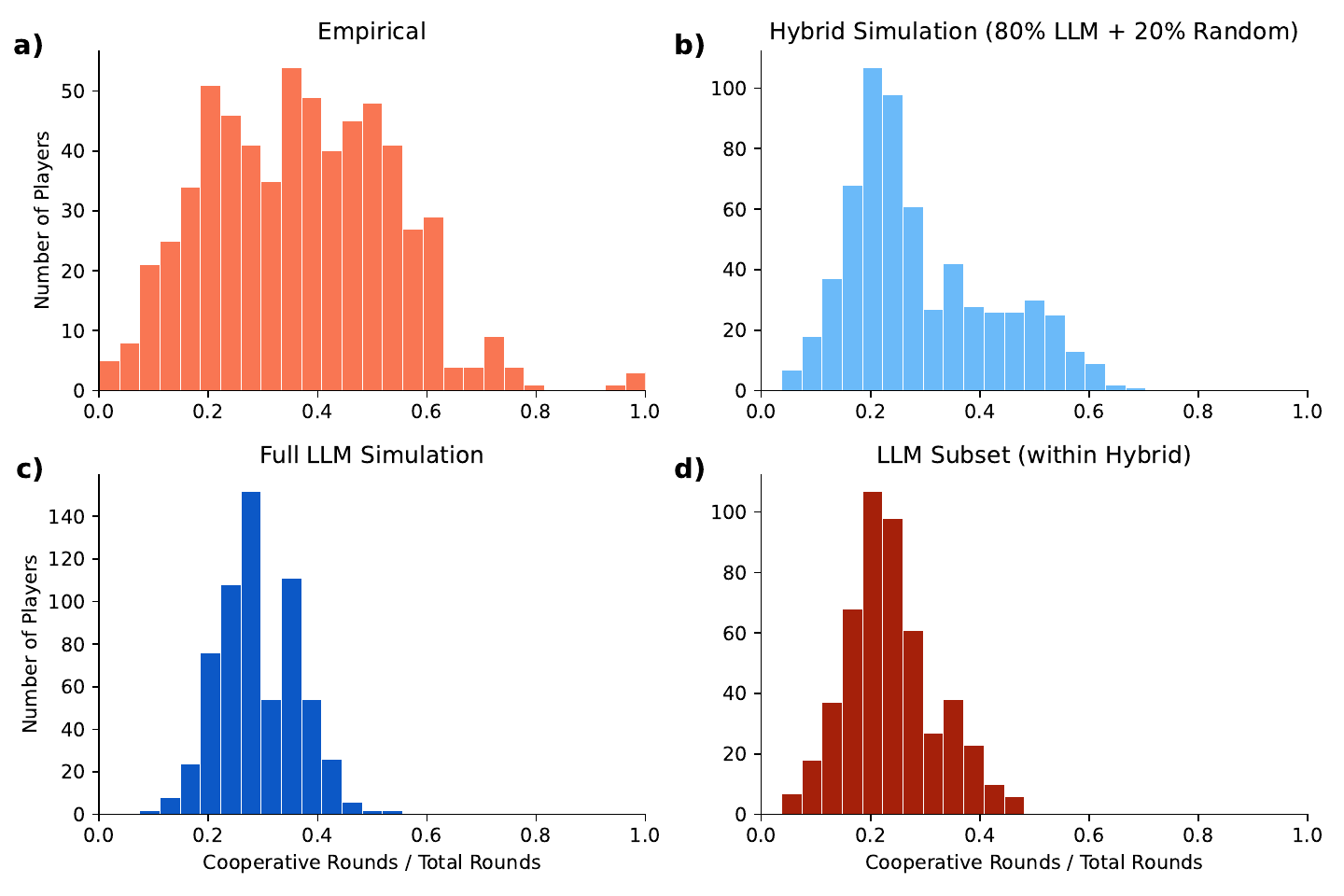}
    \caption{
    Distribution of per-agent cooperation rates for the \texttt{lattice} network with $\rho = 0.2$ (80\% LLM + 20\% Random), under the \texttt{STUDENT} condition. 
    Bins are uniformly spaced in the interval $[0,1]$. Simulation results were generated using \texttt{llama4\_16x17b}.}
    \label{fig:20_hist_lattice}
\end{figure}

\begin{figure}[h!]
    \centering
    \includegraphics[width=0.95\linewidth]{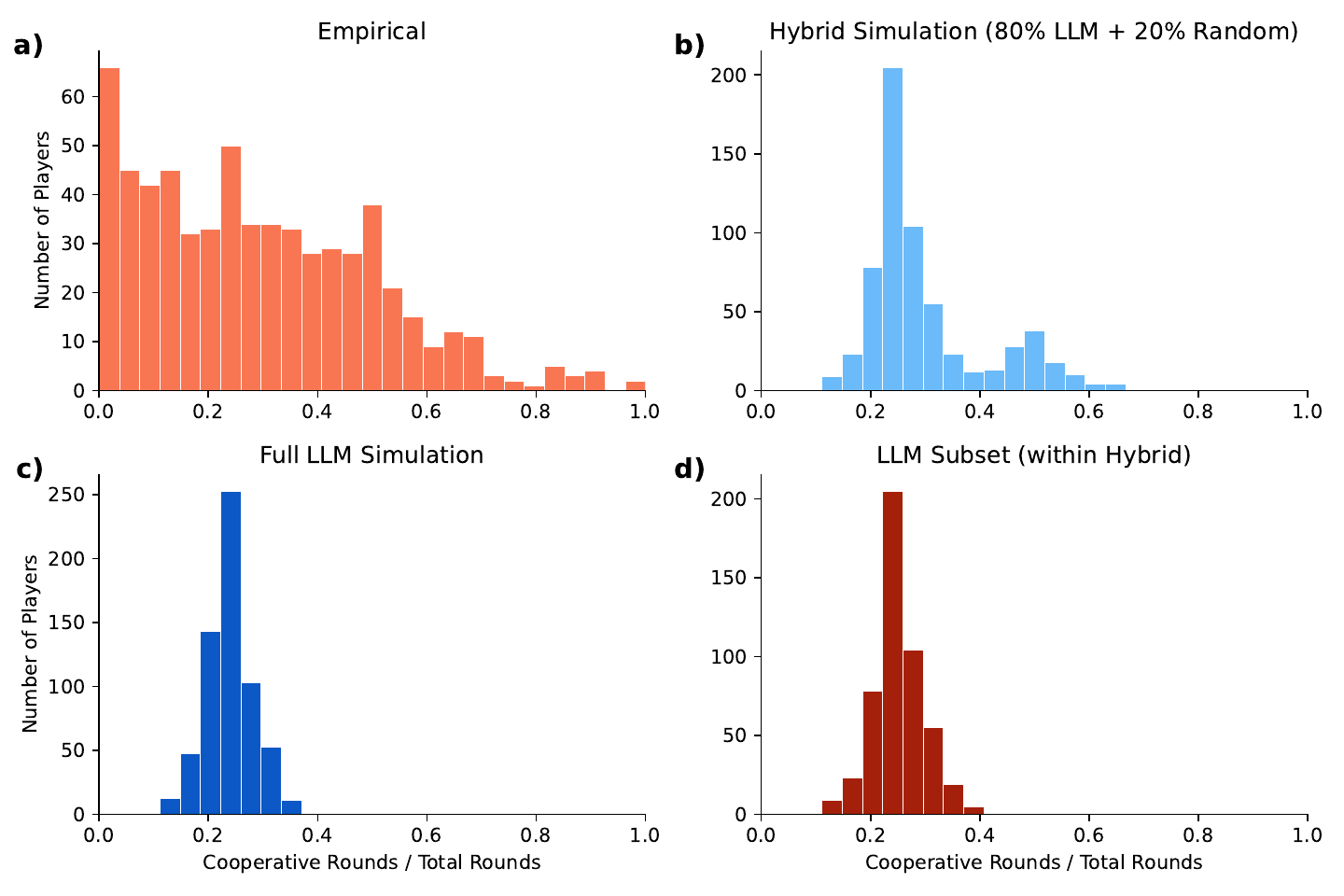}
    \caption{\textbf{Control}: Distribution of per-agent cooperation rates for the \texttt{lattice control} network with $\rho = 0.2$ (80\% LLM + 20\% Random), under the \texttt{STUDENT} condition. 
    Bins are uniformly spaced in the interval $[0,1]$. 
    Simulation results were generated using \texttt{llama4\_16x17b}.}
    \label{fig:20_hist_lattice_control}
\end{figure}

\begin{figure}[h!]
    \centering
    \includegraphics[width=0.95\linewidth]{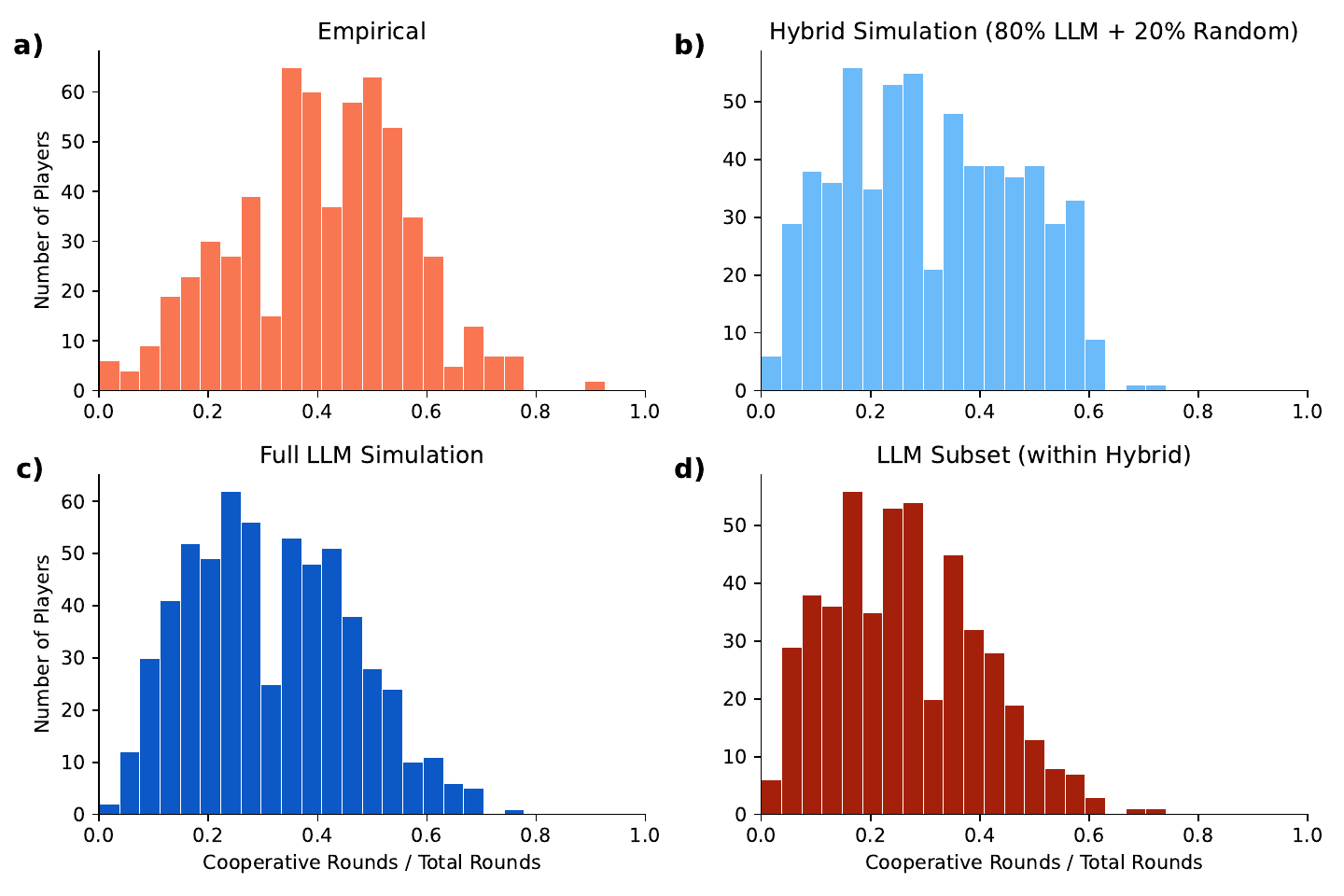}
    \caption{
    Distribution of per-agent cooperation rates for the \texttt{heterogeneous} network with $\rho = 0.2$ (80\% LLM + 20\% Random), under the \texttt{STUDENT} condition. 
    Bins are uniformly spaced in the interval $[0,1]$. 
    Simulation results were generated using \texttt{llama4\_16x17b}.}
    \label{fig:20_hist_heterogeneous}
\end{figure}

\begin{figure}[h!]
    \centering
    \includegraphics[width=0.95\linewidth]{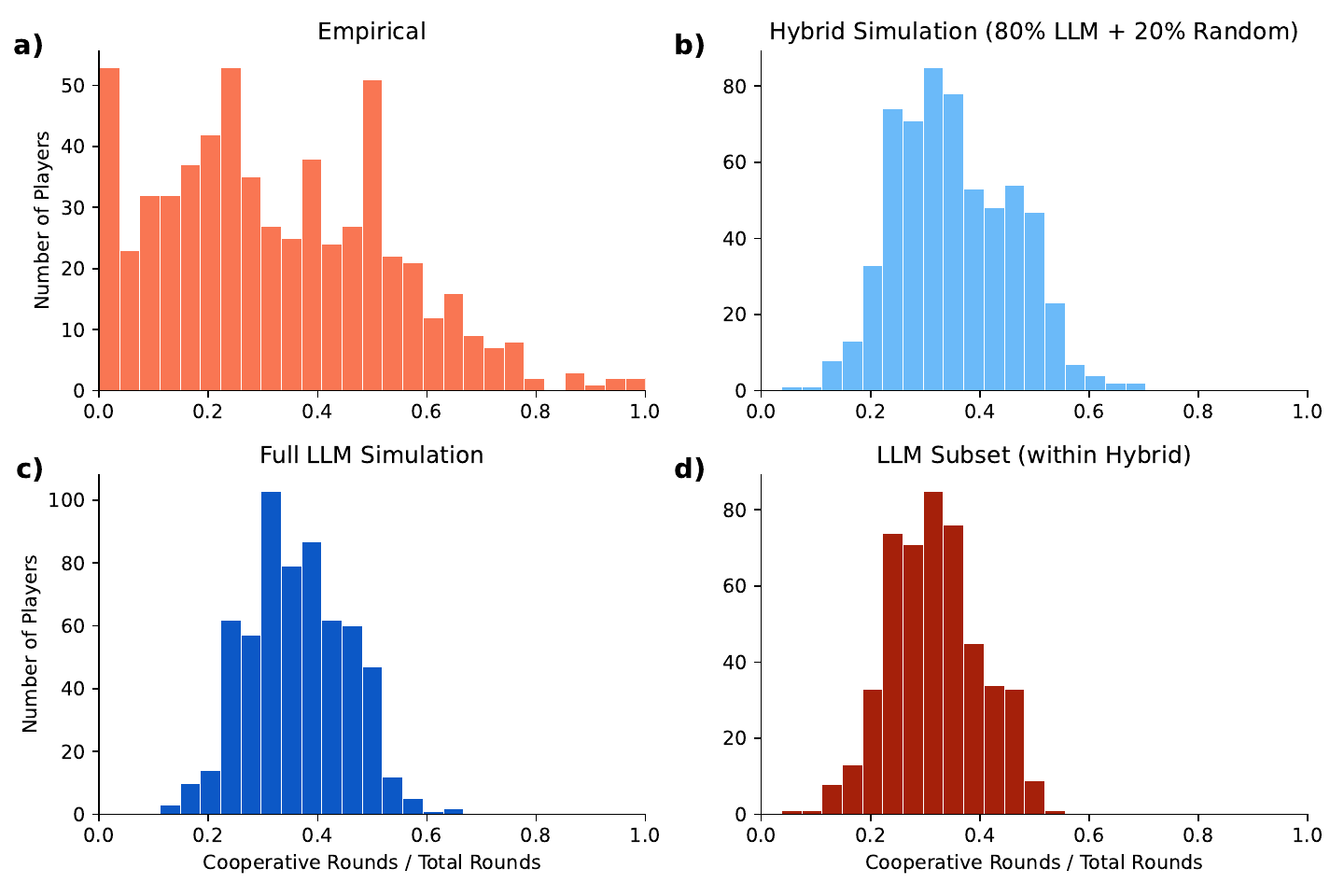}
    \caption{
    \textbf{Control}: Distribution of per-agent cooperation rates for the \texttt{heterogeneous control} network with $\rho = 0.2$ (80\% LLM + 20\% Random), under the \texttt{STUDENT} condition. 
    Bins are uniformly spaced in the interval $[0,1]$. 
    Simulation results were generated using \texttt{llama4\_16x17b}.}
    \label{fig:20_hist_heterogeneous_control}
\end{figure}
\FloatBarrier

\subsection{Quantitative analysis}

\subsubsection{Macro analysis}
\begin{figure}[h!]
    \centering
    \includegraphics[width=0.95\linewidth]{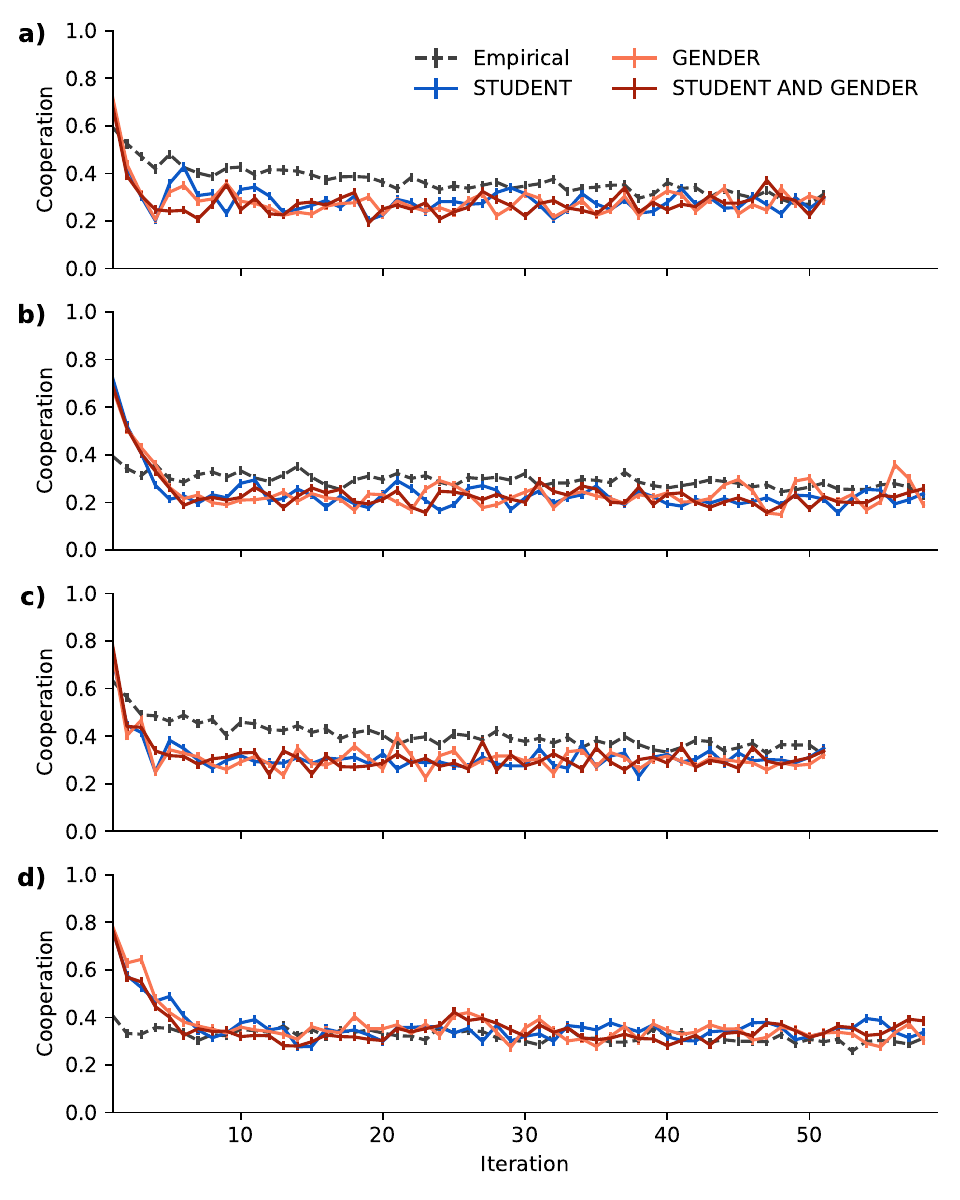}
    \caption{Evolution of cooperation over iterations for the conditions in the setting. Dashed lines show empirical results from the behavioral experiment, while solid lines show simulation results using \texttt{llama4\_16x17b}. 
    The solid lines represent the mean cooperation rate per round (empirical) or per iteration (simulation), and the error bars represent the standard error. The panels depict the following scenarios: (a) \texttt{lattice}, (b) \texttt{lattice control}, (c) \texttt{heterogeneous}, and (d) \texttt{heterogeneous control}.
    }
    \label{fig:comparison_conditions}
\end{figure}

We begin our analysis by comparing the three \texttt{backstory types}, as shown in Figure~\ref{fig:comparison_conditions} and Table~\ref{tab:Errors_types}. To reduce the number of simulations, we test two scenarios: (i) Hybrid Simulation (80\% LLM + 20\% Random) and (ii) Full LLM Simulation.

\FloatBarrier

\begin{table}[h!]
    \centering
    \scriptsize
    \begin{tabular}{lllllll}
    \toprule
    Experiment & RMSE & MAE & $\text{Diff}_\phi$ & $\text{Diff}_\mu$ & $r$ & $r^{\Delta}$ \\
    \midrule
    lattice - GENDER & 0.104** & 0.090 & 0.155*** & 0.001 & 0.577*** & 0.367*** \\
    lattice - STUDENT & 0.098** & 0.083** & 0.217*** & 0.002 & 0.612*** & 0.276* \\
    lattice - STUDENT AND GENDER & 0.110* & 0.095 & 0.113*** & 0.002 & 0.495*** & 0.384*** \\
    lattice control - GENDER & 0.089*** & 0.074* & 0.383*** & 0.003*** & 0.472*** & 0.151 \\
    lattice control - STUDENT & 0.089*** & 0.074* & 0.464*** & 0.002*** & 0.583*** & 0.214 \\
    lattice control - STUDENT AND GENDER & 0.089*** & 0.075* & 0.327*** & 0.002*** & 0.506*** & 0.031 \\
    heterogeneous - GENDER & 0.110** & 0.098 & 0.017 & 0.001* & 0.606*** & 0.117 \\
    heterogeneous - STUDENT & 0.107** & 0.096 & 0.071 & 0.002 & 0.622*** & 0.151 \\
    heterogeneous - STUDENT AND GENDER & 0.108** & 0.097 & 0.038* & 0.001* & 0.642*** & 0.224 \\
    heterogeneous control - GENDER & 0.087*** & 0.051* & 0.276 & 0.005* & 0.467*** & 0.002 \\
    heterogeneous control - STUDENT & 0.079*** & 0.051 & 0.359 & 0.004* & 0.440*** & 0.362*** \\
    heterogeneous control - STUDENT AND GENDER & 0.078*** & 0.050 & 0.420 & 0.002* & 0.367*** & 0.157 \\
    \bottomrule
    \end{tabular}

    \caption{Trajectory errors, autoregressive comparison, and correlation across different \texttt{backstory types} (for the scenario with no random agents). RMSE and MAE measure point-by-point deviations between empirical and simulated cooperation rates. $\text{Diff}_{\phi}$ denotes the absolute difference between the AR(1) persistence coefficients, and $\text{Diff}_{\mu}$ denotes the absolute difference between the implied long-run means of the simulated and empirical series. $r$ denotes the Pearson correlation coefficient between empirical and simulated trajectories. For RMSE, MAE, $\text{Diff}_{\phi}$, and $\text{Diff}_{\mu}$, statistical significance is assessed using the Monte Carlo null model described in the text. For $r$, significance refers to the null hypothesis of zero correlation ($H_0: r = 0$). $^{***}$ denotes $p < 0.01$, $^{**}$ denotes $p < 0.05$, $^{*}$ denotes $p < 0.1$, and no symbol indicates that the result is not statistically significant.}
    \label{tab:Errors_types}
\end{table}

\begin{figure}[h!]
    \centering
    \includegraphics[width=0.95\linewidth]{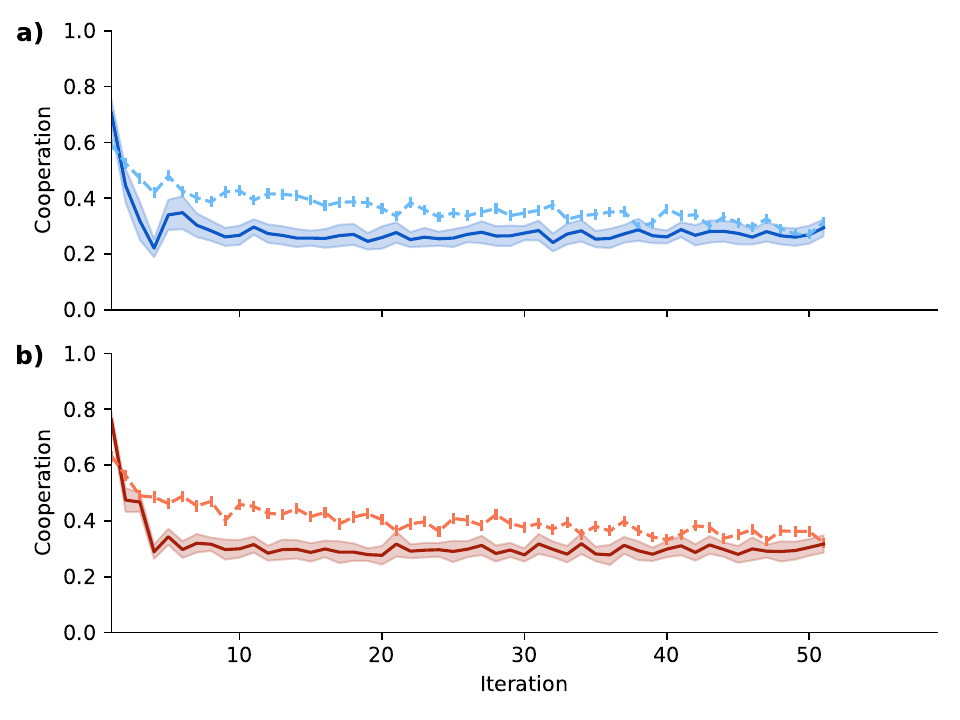}
    \caption{The average evolution of cooperation over $r=20$ runs for both lattice and heterogeneous networks (panels a and b, respectively). The dashed lines show the results of the behavioral experiment, the error bars represent the standard error, and the solid lines show the simulation results. The shaded regions represent the respective standard deviation.}
    \label{fig:comparison_runs}
\end{figure}

Since the different \texttt{backstory types} do not significantly impact the results, we only consider \texttt{STUDENT} for the following results.
First, we compare the different executions for both networks. See Figure~\ref{fig:comparison_runs} and Table~\ref{tab:Errors_runs}.

\begin{table}[h!]
    \centering
    \begin{tabular}{lllllll}
    \toprule
    Network & RMSE & MAE & $\text{Diff}_\phi$ & $\text{Diff}_\mu$ & $r$ & $r^{\Delta}$ \\
    \midrule
    Lattice & 0.099** & 0.088 & 0.461*** & 0.004** & 0.666*** & 0.407*** \\
    Heterogeneous & 0.105** & 0.096* & 0.190 & 0.001** & 0.708*** & 0.179 \\
    \bottomrule
    \end{tabular}
    
    \caption{Trajectory errors, autoregressive comparison, and correlation across different executions for the scenario with no random agents, on both \texttt{lattice} and \texttt{heterogeneous} networks. RMSE and MAE measure point-by-point deviations between empirical and average simulated cooperation rates. $\text{Diff}_{\phi}$ denotes the absolute difference between the AR(1) persistence coefficients, and $\text{Diff}_{\mu}$ denotes the absolute difference between the implied long-run means of the simulated and empirical series. $r$ denotes the Pearson correlation coefficient between empirical and simulated trajectories. For RMSE, MAE, $\text{Diff}_{\phi}$, and $\text{Diff}_{\mu}$, statistical significance is assessed using the Monte Carlo null model described in the text. For $r$, significance refers to the null hypothesis of zero correlation ($H_0: r = 0$). $^{***}$ denotes $p < 0.01$, $^{**}$ denotes $p < 0.05$, $^{*}$ denotes $p < 0.1$, and no symbol indicates that the result is not statistically significant.}
    \label{tab:Errors_runs}
\end{table}
\FloatBarrier

Table~\ref{tab:Errors_percentage_rand} reports trajectory errors, AR(1) differences, and correlation metrics for simulations on \texttt{lattice} network, with varying percentages of random agents in the population. Table~\ref{tab:trajectory_metrics} presents the same metrics across different experimental scenarios (i.e., lattice, heterogeneous, and control setups).


\begin{table}[h!]
    \centering
    \begin{tabular}{rllllll}
    \toprule
    Random (\%) & RMSE & MAE & $\text{Diff}_\phi$ & $\text{Diff}_\mu$ & $r$ & $r^{\Delta}$ \\
    \midrule
    0 & 0.098** & 0.083** & 0.217*** & 0.002 & 0.612*** & 0.276* \\
    10 & 0.116*** & 0.099** & 0.537*** & 0.000** & 0.665*** & 0.284** \\
    20 & 0.109*** & 0.093** & 0.405*** & 0.001*** & 0.690*** & 0.469*** \\
    30 & 0.104*** & 0.092** & 0.443*** & 0.000*** & 0.691*** & 0.384*** \\
    40 & 0.086* & 0.074 & 0.098*** & 0.001 & 0.675*** & 0.282** \\
    50 & 0.057*** & 0.044** & 0.376*** & 0.003 & 0.638*** & 0.200 \\
    60 & 0.049** & 0.037** & 0.020*** & 0.003 & 0.569*** & 0.236* \\
    70 & 0.060*** & 0.048*** & 0.058* & 0.003 & 0.645*** & 0.332** \\
    80 & 0.091 & 0.079 & 0.048* & 0.005*** & 0.304** & -0.097 \\
    90 & 0.121 & 0.110 & 0.215* & 0.004** & 0.457*** & 0.297** \\
    \bottomrule
    \end{tabular}

    \caption{Trajectory errors, autoregressive comparison, and correlation across different percentages of random agents for the same \texttt{backstory type} on \texttt{lattice} network. RMSE and MAE measure point-by-point deviations between empirical and simulated cooperation rates. $\text{Diff}_{\phi}$ denotes the absolute difference between the AR(1) persistence coefficients, and $\text{Diff}_{\mu}$ denotes the absolute difference between the implied long-run means of the simulated and empirical series. $r$ denotes the Pearson correlation coefficient between empirical and simulated trajectories. For RMSE, MAE, $\text{Diff}_{\phi}$, and $\text{Diff}_{\mu}$, statistical significance is assessed using the Monte Carlo null model described in the text. For $r$, significance refers to the null hypothesis of zero correlation ($H_0: r = 0$). $^{***}$ denotes $p < 0.01$, $^{**}$ denotes $p < 0.05$, $^{*}$ denotes $p < 0.1$, and no symbol indicates that the result is not statistically significant.}
    \label{tab:Errors_percentage_rand}
\end{table}

\begin{table}[h!]
    \centering
    \scriptsize
    \begin{tabular}{lllllll}
    \toprule
    Experiment & RMSE & MAE & $\text{Diff}_\phi$ & $\text{Diff}_\mu$ & $r$ & $r^{\Delta}$ \\
    \midrule
    lattice (100\% LLM + 0\% Rand.) & 0.098** & 0.083** & 0.217*** & 0.002 & 0.612*** & 0.276* \\
    lattice (80\% LLM + 20\% Rand.) & 0.109*** & 0.093** & 0.405*** & 0.001*** & 0.690*** & 0.469*** \\
    lattice control (100\% LLM + 0\% Rand.) & 0.089*** & 0.074* & 0.464** & 0.002*** & 0.583*** & 0.214 \\
    lattice control (80\% LLM + 20\% Rand.) & 0.087*** & 0.046** & 0.519*** & 0.005*** & 0.509*** & 0.098 \\
    heterogeneous (100\% LLM + 0\% Rand.) & 0.107** & 0.096 & 0.071 & 0.002 & 0.622*** & 0.151 \\
    heterogeneous (80\% LLM + 20\% Rand.) & 0.109*** & 0.101** & 0.249 & 0.000*** & 0.743*** & 0.343** \\
    heterogeneous control (100\% LLM + 0\% Rand.) & 0.079*** & 0.051 & 0.359 & 0.004* & 0.440*** & 0.362*** \\
    heterogeneous control (80\% LLM + 20\% Rand.) & 0.070** & 0.046 & 0.384 & 0.003* & 0.325** & 0.135 \\
    \bottomrule
    \end{tabular}
    
    \caption{Trajectory errors, autoregressive comparison, and correlation across experimental scenarios. RMSE and MAE measure point-by-point deviations between empirical and simulated cooperation rates. $\text{Diff}_{\phi}$ denotes the absolute difference between the AR(1) persistence coefficients, and $\text{Diff}_{\mu}$ denotes the absolute difference between the implied long-run means of the simulated and empirical series. $r$ denotes the Pearson correlation coefficient between empirical and simulated trajectories. For RMSE, MAE, $\text{Diff}_{\phi}$, and $\text{Diff}_{\mu}$, statistical significance is assessed using the Monte Carlo null model described in the text. For $r$, significance refers to the null hypothesis of zero correlation ($H_0: r = 0$). $^{***}$ denotes $p < 0.01$, $^{**}$ denotes $p < 0.05$, $^{*}$ denotes $p < 0.1$, and no symbol indicates that the result is not statistically significant.}
\label{tab:trajectory_metrics}
\end{table}
\FloatBarrier

\subsubsection{Micro analysis}
First, we show the persistence across different percentages of random agents (in the \texttt{lattice} scenario), as shown in Table~\ref{tab:mean_coop_ci_rand}. Next, we present the comparison between the different experiments (see Table~\ref{tab:mean_coop_ci}).

\begin{table}[h!]
    \centering
    \begin{tabular}{lll}
    \toprule
    Experiment & $\overline{\mathcal{P}}^{\mathcal{D}}$ & 95\% CI \\
    \midrule
    Empirical & 0.640 & (0.629, 0.652) \\
    Lattice (100\% LLM + 0\% Rand.) & 0.579 & (0.573, 0.586) \\
    Lattice (90\% LLM + 10\% Rand.) & 0.613 & (0.605, 0.621) \\
    Lattice (80\% LLM + 20\% Rand.) & 0.633 & (0.624, 0.642) \\
    Lattice (70\% LLM + 30\% Rand.) & 0.645 & (0.634, 0.656) \\
    Lattice (60\% LLM + 40\% Rand.) & 0.650 & (0.637, 0.662) \\
    Lattice (50\% LLM + 50\% Rand.) & 0.610 & (0.598, 0.621) \\
    Lattice (40\% LLM + 60\% Rand.) & 0.593 & (0.582, 0.605) \\
    Lattice (30\% LLM + 70\% Rand.) & 0.564 & (0.554, 0.574) \\
    Lattice (20\% LLM + 80\% Rand.) & 0.540 & (0.531, 0.549) \\
    Lattice (10\% LLM + 90\% Rand.) & 0.518 & (0.511, 0.525) \\
    \bottomrule
    \end{tabular}

    \caption{Mean cooperation persistence and 95\% confidence interval for different percentages of random agents. Confidence intervals are estimated via bootstrapping over agents.}
    \label{tab:mean_coop_ci_rand}
\end{table}

\begin{table}[h!]
    \centering
    \begin{tabular}{lll}
    \toprule
    Experiment & $\overline{\mathcal{P}}^{\mathcal{D}}$ & 95\% CI \\
    \midrule
    lattice (100\% LLM + 0\% Rand.) & 0.579 & (0.573, 0.586) \\
    lattice (80\% LLM + 20\% Rand.) & 0.633 & (0.623, 0.642) \\
    lattice - Empirical & 0.640 & (0.629, 0.651) \\
    \midrule
    lattice control (100\% LLM + 0\% Rand.) & 0.610 & (0.604, 0.615) \\
    lattice control (80\% LLM + 20\% Rand.) & 0.589 & (0.582, 0.596) \\
    lattice control - Empirical & 0.724 & (0.711, 0.738) \\
    \midrule
    heterogeneous (100\% LLM + 0\% Rand.) & 0.604 & (0.592, 0.617) \\
    heterogeneous (80\% LLM + 20\% Rand.) & 0.629 & (0.615, 0.642) \\
    heterogeneous - Empirical & 0.620 & (0.610, 0.630) \\
    \midrule
    heterogeneous control (100\% LLM + 0\% Rand.) & 0.578 & (0.572, 0.584) \\
    heterogeneous control (80\% LLM + 20\% Rand.) & 0.577 & (0.570, 0.584) \\
    heterogeneous control - Empirical & 0.693 & (0.680, 0.707) \\
    \bottomrule
    \end{tabular}

    \caption{Mean cooperation persistence and 95\% confidence interval. Confidence intervals are estimated via bootstrapping over agents.}
    \label{tab:mean_coop_ci}
\end{table}
\FloatBarrier

\begin{figure}[h!]
\centering
\includegraphics[width=0.9\linewidth]{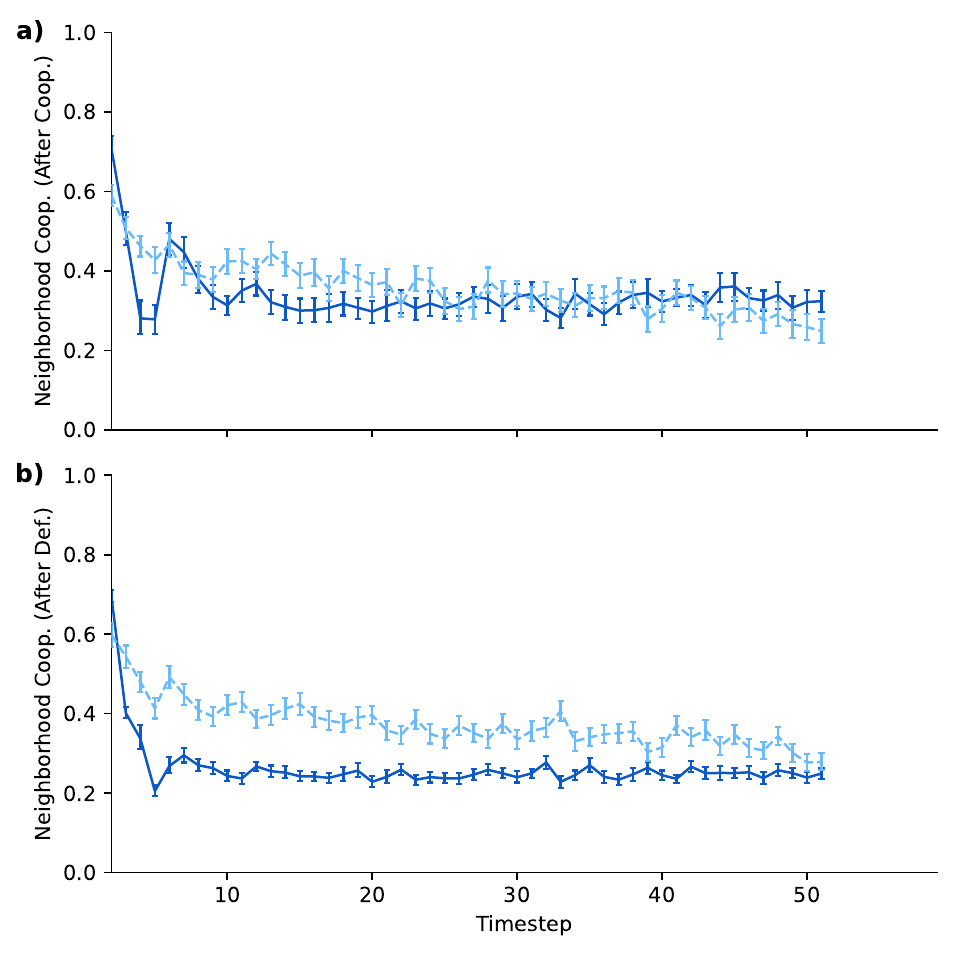}
\caption{Average fraction of cooperating neighbors over time for the \texttt{lattice} network. Results are shown separately for agents that cooperated (panel a) or defected (panel b) in the previous round. The continuous blue line represents the simulation, and the black dashed line represents the empirical data. The simulated curve is the mean across all 20 runs, and the error bars represent 95\% CIs estimated by Cluster bootstrap across runs. Empirical patterns (dashed) are shown with 95\% CIs estimated by bootstrapping over players.}
\label{fig:neighborhood_cooperation_same_dynamics_lattice}
\end{figure}

\begin{figure}[h!]
\centering
\includegraphics[width=0.9\linewidth]{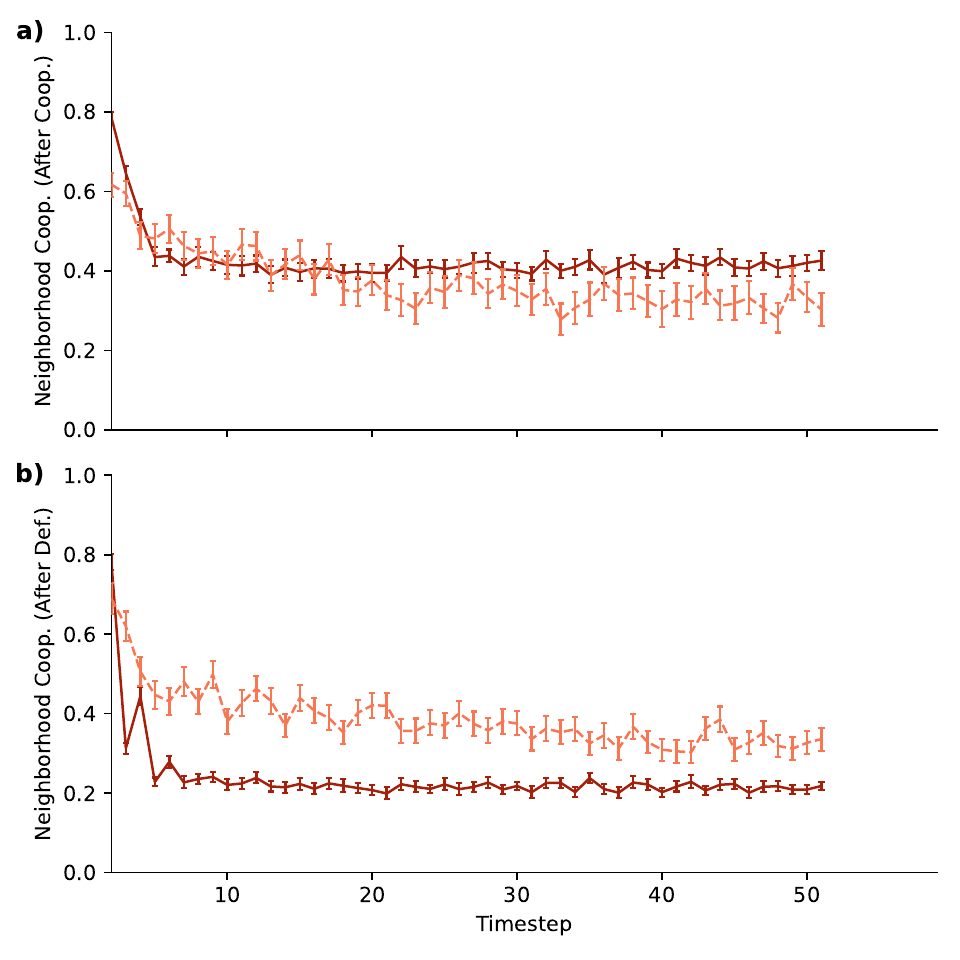}
\caption{Average fraction of cooperating neighbors over time for the \texttt{heterogeneous} network. Results are shown separately for agents that cooperated (panel a) or defected (panel b) in the previous round. The continuous green line represents the simulation, and the black dashed line represents the empirical data. The simulated curve is the mean across all 20 runs, and the error bars represent 95\% CIs estimated by Cluster bootstrap across runs. Empirical patterns (dashed) are shown with 95\% CIs estimated by bootstrapping over players.}
\label{fig:neighborhood_cooperation_same_dynamics_heterogeneous}
\end{figure}

Next, we examine conditional cooperation as a function of neighborhood behavior. Figures~\ref{fig:neighborhood_cooperation_same_dynamics_lattice}~and~\ref{fig:neighborhood_cooperation_same_dynamics_heterogeneous} show the average probability of cooperation as a function of the fraction of cooperating neighbors in the previous round, separately after a previous defection and after a previous cooperation. Both tested scenarios (\texttt{lattice} and \texttt{heterogeneous}) comprise 20 dynamics executions.

\FloatBarrier

\subsubsection{Heterogeneity across individuals}
To further analyze the role of stochastic behavior in shaping individual-level heterogeneity, we vary the fraction of random agents $\rho$ in the population. Table~\ref{tab:heterogeneity_results_rand} reports the resulting values of $W_1(E,S)$ across different percentages of random agents. The results show a non-monotonic relationship between $\rho$ and heterogeneity mismatch, with intermediate levels of randomness generally yielding lower distances between empirical and simulated distributions.

\begin{table}[h!]
\centering
\begin{tabular}{rl}
    \toprule
    Random (\%) & $W_1(E,S)$ \\
    \midrule
    0 & 0.101 \\
    10 & 0.089 \\
    20 & 0.081 \\
    30 & 0.084 \\
    40 & 0.073 \\
    50 & 0.037 \\
    60 & 0.039 \\
    70 & 0.051 \\
    80 & 0.080 \\
    90 & 0.119 \\
    \bottomrule
\end{tabular}

\caption{Sensitivity of individual cooperation heterogeneity to the fraction of random agents. Wasserstein distance between empirical and simulated distributions of individual cooperation propensities. Lower values indicate better reproduction of behavioral heterogeneity.}
\label{tab:heterogeneity_results_rand}
\end{table}
\FloatBarrier

\subsubsection{Conditional cooperation rule}
To quantify how well the simulations reproduce the empirical conditional cooperation rule, we compute RMSE and MAE between the empirical and simulated conditional cooperation probabilities. Errors are calculated separately for cases in which the previous action was defection ($z=0$) and cooperation ($z=1$), as well as aggregated across both cases, see Table~\ref{tab:table_conditional_rule_percentages_rand}.

\begin{table}[h!]
\centering
\begin{tabular}{rllll}
\toprule
Random (\%) & RMSE$_{z=0}$ & MAE$_{z=0}$ & RMSE$_{z=1}$ & MAE$_{z=1}$ \\ 
\midrule
0 & 0.237** & 0.217*** & 0.397*** & 0.388* \\
10 & 0.208** & 0.188*** & 0.322*** & 0.310* \\
20 & 0.180 & 0.165*** & 0.255*** & 0.246* \\
30 & 0.153 & 0.144*** & 0.199*** & 0.192* \\
40 & 0.145** & 0.136*** & 0.117*** & 0.116*** \\
50 & 0.140*** & 0.121*** & 0.096*** & 0.092*** \\
60 & 0.137*** & 0.112*** & 0.060*** & 0.055*** \\
70 & 0.135*** & 0.101*** & 0.049*** & 0.041*** \\
80 & 0.150*** & 0.116*** & 0.042*** & 0.038*** \\
90 & 0.181*** & 0.163*** & 0.035*** & 0.027*** \\
\bottomrule
\end{tabular}

\caption{Deviation between empirical and simulated conditional cooperation rules across different percentages of random agents. 
Errors are computed separately for cases in which the previous action was defection ($z=0$) or cooperation ($z=1$). 
The columns RMSE$_{all}$ and MAE$_{all}$ summarize the overall deviation across both cases. $^{***}$ denotes $p < 0.01$, $^{**}$ denotes $p < 0.05$, $^{*}$ denotes $p < 0.1$, and no symbol indicates that the result is not statistically significant.}
\label{tab:table_conditional_rule_percentages_rand}
\end{table}
\FloatBarrier


\clearpage
\bibliography{refs}

\end{document}